\documentclass[12pt,a4paper]{article}
\usepackage[T1]{fontenc}
\usepackage[utf8]{inputenc}
\usepackage{lmodern,microtype}
\usepackage[margin=25mm]{geometry}
\usepackage{amsmath,amssymb,amsthm,mathtools,bm}
\usepackage{graphicx,tikz,pgfplots,booktabs,array}
\pgfplotsset{compat=1.16}
\usepackage{xcolor}
\usepackage[colorlinks=true,linkcolor=blue!45!black,citecolor=blue!45!black,urlcolor=blue!45!black]{hyperref}
\hypersetup{pdftitle={Geometry of squeezed states: the phase-space metric, relative vacua and cosmology},pdfauthor={Alejandro Corichi and Jer\'onimo Cortez}}
\definecolor{deepblue}{RGB}{28,64,93}
\definecolor{teal}{RGB}{16,120,125}
\definecolor{rust}{RGB}{177,77,46}
\numberwithin{equation}{section}
\theoremstyle{plain}
\newtheorem{theorem}{Theorem}[section]
\newtheorem{proposition}[theorem]{Proposition}
\newtheorem{corollary}[theorem]{Corollary}
\theoremstyle{definition}
\newtheorem{definition}[theorem]{Definition}
\newcommand{\Om}{\Omega}
\newcommand{\Id}{\mathbf{1}}
\newcommand{\Tr}{\operatorname{Tr}}
\newcommand{\Cov}{\operatorname{Cov}}
\newcommand{\Sp}{\operatorname{Sp}}
\newcommand{\diag}{\operatorname{diag}}
\newcommand{\spec}{\operatorname{spec}}
\newcommand{\dd}{\mathrm{d}}
\newcommand{\vac}[1]{\lvert 0;#1\rangle}
\newcommand{\R}{\mathbb{R}}

\allowdisplaybreaks[1]

\title{\textbf{Geometry of squeezed states}\\[5pt]\large The phase-space metric, relative vacua and cosmology}
\author{Alejandro Corichi$^{1}$ and Jer\'onimo Cortez$^{2}$\\[6pt]
\small $^{1}$Centro de Ciencias Matem\'aticas, Universidad Nacional Aut\'onoma de M\'exico,\\ \small UNAM-Campus Morelia, A. Postal 61-3, Morelia, Michoac\'an 58190, Mexico \\ \small{\tt{corichi@matmor.unam.mx}}\\[2pt]
\small $^{2}$Departamento de F\'isica, Facultad de Ciencias, Universidad Nacional Aut\'onoma de M\'exico,\\ \small Ciudad de M\'exico 04510, Mexico \\ \small{\tt{jacq@ciencias.unam.mx}}}
\date{}

\begin{document}
\maketitle

\begin{abstract}
Geometry enters quantum theory in two places. The space of quantum states carries a symplectic form, which generates unitary evolution, and a metric, which governs uncertainties and transition probabilities. The classical phase space acquires a metric of its own when a vacuum is chosen: quantizing a linear system amounts to choosing a compatible complex structure, equivalently a compatible metric $g$. We take this second metric as the primary object and show that it plays, on phase space, the role that the quantum metric plays on the space of states: fluctuations of linear observables are $g$-norms of Hamiltonian vector fields, and along coherent states $g/2\hbar$ \emph{is} the quantum metric. Squeezed states then have a simple description. A squeezed vacuum is the vacuum of another metric, and squeezing is a property of a pair of metrics, a relative vacuum, whose invariants are the eigenvalues $e^{\pm2r_i}$ of $L=g_0^{-1}g$. We work out the harmonic oscillator in detail, including why a rotated squeezed state no longer minimizes the product of the displayed uncertainties. Quadratic evolution deforms the metric, and only the part of the generator that is self-adjoint with respect to the reference changes the $r_i$, at a rate we give exactly. For fields, a vacuum is unitarily implementable relative to a reference exactly when it lies at finite distance from it in the space of vacua. For a massless field in de Sitter space we revisit the two canonical charts compared by Grain and Vennin: relative to the reference of the standard squeezing formalism the Bunch--Davies state has $\sinh r_k=1/(2|k\eta|)$, and relative to the vacuum that a radiation era beginning at time $\eta$ would select it has $\sinh r_k=1/(2k^2\eta^2)$. The second comparison squeezes twice as fast and is unitarily implementable; the first is not.
\end{abstract}

\noindent\textbf{Keywords:} squeezed states, geometric quantum mechanics, K\"ahler structures, Gaussian states, Fubini--Study metric, quantum fields in curved spacetime, cosmological perturbations

\section{Introduction}
\label{sec:intro}

There are two different ways in which geometric objects are fundamental in quantum theory. The first is the geometry of the space of states. The Hermitian inner product of a Hilbert space splits into a symplectic form and a Riemannian metric, both of which descend to the projective space of rays \cite{kibble,ashsch}. The symplectic form generates unitary evolution, which becomes Hamiltonian flow on the space of states. The metric is responsible for the more genuinely quantum features: it measures uncertainties and transition probabilities. The second place is quantization itself. The classical phase space $\Gamma$ of a linear system is already a symplectic space, but to quantize it one must choose a compatible complex structure $J$, equivalently a compatible metric $g$ \cite{am}. Everything else, the one-particle Hilbert space, the Fock space, and the preferred vacuum, follows from that choice. The phase space then carries the same kind of structure, a symplectic form together with a metric, as the space of quantum states.

This paper is about the second metric and what it teaches about squeezed states, which have long been described through wavefunctions, uncertainty ellipses and Bogoliubov transformations \cite{stoler,yuen}. Three questions guide it. What physical information does the phase-space metric $g$ carry? Can it be used to understand and visualize squeezed states? And is it useful in cosmology, where squeezing of primordial perturbations is central to discussions of their apparent classicality \cite{gs,afjp,ps,mv,ack}? The questions grew out of the analysis of classical behavior in the early universe \cite{ack}, a talk at the BIRS--CMO workshop on mathematical and conceptual aspects of quantum theory \cite{cmo}, and our working notes on the harmonic oscillator.

The answers fit a single picture. On the space of states the uncertainty of an observable is the norm of the vector field it generates; on phase space the variance of a linear observable $f$ is $\frac\hbar2 g(X_f,X_f)$, the norm of its Hamiltonian vector field. Along coherent states the two statements coincide: $g/2\hbar$ is exactly the Fubini--Study metric (Section~\ref{sec:geometry}). A vacuum is thus a choice of metric on phase space, and a squeezed vacuum is the vacuum of another metric. ``Squeezed'' is always relative: a state is squeezed only with respect to a reference vacuum, and squeezing is a property of a \emph{pair} of metrics, which we call a relative vacuum (Section~\ref{sec:squeezed}). The pair has exactly $N$ invariants, the relative squeezing parameters $r_i$, and every coordinate-independent comparison of two vacua is a function of them.

\paragraph{What is known.} Much of the mathematics is standard. Pure Gaussian states are described by covariance matrices, compatible complex structures and relative complex structures \cite{smd,arvind,bhy,hb}. The normal form of a pair of vacua is the Bloch--Messiah, or Cartan, decomposition \cite{arvind,braunstein}. The quantum metric of coherent states is the K\"ahler metric of phase space \cite{kibble,provost,perelomov,ashsch}, and the Fubini--Study metric of the Gaussian family appears in Gaussian quantum estimation \cite{monras,safranek}. Unitary implementability is governed by Shale's theorem \cite{shale,dg}. For cosmological perturbations, Grain and Vennin \cite{gv} showed that different canonical variables select different vacua. For a test field in de Sitter space they compared exactly the two canonical charts used in Section~\ref{sec:cosmology}, computed the squeeze relating the two reference vacua and the leading-order Bogoliubov coefficients of the Bunch--Davies state relative to each, and showed that the vacuum of the first chart has zero overlap with the Bunch--Davies state. Agullo, Bonga and Ribes Metidieri \cite{abrm} argued that in a generic FLRW background the squeezing of a pair of modes $(\bm k,-\bm k)$ is as ambiguous as particle creation, and that once the symmetries of de Sitter space are used to resolve the ambiguity, inflation does not squeeze the perturbations.

\paragraph{What this paper adds.} We organize this material around the phase-space metric, and within that framework obtain, to our knowledge for the first time in this form:
\begin{enumerate}
\item an exact formula for the rate of change of the squeezing parameters under quadratic evolution, in which only the part of the generator that is self-adjoint with respect to the reference contributes, together with a bound (Proposition~\ref{prop:rate});
\item the equivalence of unitary implementability, finite reference particle number, nonvanishing vacuum overlap, and finite distance in the space of Gaussian vacua (Theorem~\ref{thm:shale});
\item for the two de Sitter references of Ref.~\cite{gv}: exact expressions for the squeezing of the Bunch--Davies state at all times; the late-time rates, one and two units per $e$-fold, and their origin in the squeezing part of the generator; the identification of the second reference with the vacuum of a subsequent radiation era; and the statement that only the second comparison is unitarily implementable, both at fixed time and for the dynamics, the latter through two-sided bounds on the Bogoliubov coefficients of the evolution that follow from the fixed-time comparison (Sections~\ref{sec:cosmology} and~\ref{sec:fields}, Proposition~\ref{prop:dynamics}).
\end{enumerate}
Along the way we separate three operations that are often conflated, passive re-description, re-prescription of the vacuum, and active squeezing (Figures~\ref{fig:ellipses} and~\ref{fig:disk}), and we resolve an apparent puzzle about rotated squeezed states (Section~\ref{sec:rotation}).

Two companion papers build on this framework. Reference~\cite{ls} asks which reference the dynamics itself suggests, and studies the vacua that minimize the squeezing part of the generator. Reference~\cite{prd} fixes a reference and quantifies how far the reported cosmological squeezing can change under bounded changes of the state and of the reference.

We work with finitely many canonical pairs and centered pure Gaussian states until Section~\ref{sec:fields}, where unitary equivalence becomes a substantive question. The paper is organized as follows. Section~\ref{sec:geometry} sets up the two geometries and the formula they share. Section~\ref{sec:oscillator} treats the harmonic oscillator in detail. Section~\ref{sec:squeezed} describes squeezed vacua and relative vacua. Section~\ref{sec:dynamics} treats evolution, Section~\ref{sec:cosmology} the cosmological example, Section~\ref{sec:rays} the geometry of the space of vacua, and Section~\ref{sec:fields} quantum fields. Section~\ref{sec:discussion} discusses the results and Section~\ref{sec:conclusions} concludes.

\section{Two geometries, one formula}
\label{sec:geometry}

\subsection{The geometry of quantum states}

A Hilbert space $\mathcal H$, regarded as a real vector space, carries two structures read off from its Hermitian inner product: its real part defines a Riemannian metric $G$ and its imaginary part a symplectic form $\Om_{\mathcal H}$. Both descend to the projective space $\mathcal P$ of rays, which is the true space of pure states \cite{kibble,ashsch}. Unitary evolution generated by $\hat H$ becomes, on $\mathcal P$, the Hamiltonian flow of the expectation-value function $\langle\hat H\rangle$ with respect to $\Om_{\mathcal H}$: Schr\"odinger's equation is Hamilton's equation, and unitary transformations are symplectomorphisms that also preserve $G$. The metric carries the probabilistic content. In the dimensionless Fubini--Study normalization that we use,
\begin{equation}
 \dd s_{\rm FS}^2=\langle\dd\psi|\dd\psi\rangle-|\langle\psi|\dd\psi\rangle|^2,
 \qquad \langle\psi|\psi\rangle=1,
 \label{eq:FS}
\end{equation}
the transition probability between two rays is $\cos^2$ of their geodesic distance, and the uncertainty of an observable $\hat A$ is the speed at which it moves the state \cite{aa}:
\begin{equation}
 (\Delta\hat A)^2=\hbar^2\left(\frac{\dd s_{\rm FS}}{\dd t}\right)^2
 \quad\text{along}\quad \psi(t)=e^{-it\hat A/\hbar}\psi .
 \label{eq:AAspeed}
\end{equation}
In the normalization of Ref.~\cite{ashsch}, this reads $(\Delta\hat A)^2=\frac\hbar2G(X_A,X_A)$, with $X_A$ the Hamiltonian vector field of $\langle\hat A\rangle$. Uncertainty is the metric norm of a Hamiltonian vector field.

\subsection{The geometry of a quantized phase space}

Let $\Gamma\simeq\R^{2N}$ be a linear phase space. In coordinates $x=(q^1,\ldots,q^N,p_1,\ldots,p_N)^T$, choose
\begin{equation}
 \Om=\sum_i\dd p_i\wedge\dd q^i,
 \qquad \Pi:=\Om^{-1}=\begin{pmatrix}0&\Id\\-\Id&0\end{pmatrix},
 \label{eq:omega}
\end{equation}
so that $\Om\bigl((q_1,p_1),(q_2,p_2)\bigr)=p_1\cdot q_2-p_2\cdot q_1$. The same symbol denotes a bilinear form and its matrix when no confusion is possible. Hamiltonian vector fields satisfy
\begin{equation}
 \dd f=\Om(\,\cdot\,,X_f),\qquad X_f=\Pi\,\dd f,
 \qquad \{f,h\}=(\dd f)^T\Pi\,\dd h,
\end{equation}
so $X_q=-\partial_p$, $X_p=\partial_q$, $\{q,p\}=1$, and quantization gives $[\hat q,\hat p]=i\hbar$. Quantizing a linear system requires one more choice, a compatible complex structure, that is, a real linear map $J$ with
\begin{equation}
 J^2=-\Id,\qquad J^T\Om J=\Om,
 \qquad g_J(u,v):=\Om(Ju,v),\quad g_J(u,u)>0\ \ (u\ne0).
 \label{eq:compat}
\end{equation}
In matrices\footnote{Refs.~\cite{ack,corichicl} write $g(u,v)=\Om(u,Jv)$ and use the opposite sign for the matrix of $\Om$. The two sign changes compensate: for a given state, the matrices of $J$ and $g$ coincide with ours.}
\begin{equation}
 g_J=J^T\Om=-\Om J,\qquad J=-\Pi g_J,
 \qquad g_J\Pi g_J=-\Om.
 \label{eq:matrixmetric}
\end{equation}
The last equality is the compatibility condition on the metric; an arbitrary positive metric does not satisfy it. Once $\Om$ is fixed, a compatible metric and a compatible complex structure carry the same information. Together with $\Om$, the metric $g_J$ defines the Hermitian product $\langle\!\langle\gamma_1,\gamma_2\rangle\!\rangle=\frac12g_J(\gamma_1,\gamma_2)-\frac i2\Om(\gamma_1,\gamma_2)$ on $\Gamma$, regarded as a complex vector space through $J$; this is the one-particle Hilbert space, and $(\Gamma,\Om,g_J)$ has the same structure as $(\mathcal P,\Om_{\mathcal H},G)$. The $-i$ eigenspace of $J$ defines the annihilation variables (Section~\ref{sec:modes}), and their common kernel is the vacuum ray, which we denote $\vac{J}$. All such vacua can be represented in a common irreducible representation when $N$ is finite \cite{bhy,hb}.

\subsection{Fluctuations as metric norms}

For real linear observables $f$ and $h$, the symmetrized covariance in the vacuum is
\begin{equation}
 \Cov_J(f,h):=\tfrac12\langle\hat f\hat h+\hat h\hat f\rangle_J
 -\langle\hat f\rangle_J\langle\hat h\rangle_J
 =\frac{\hbar}{2}\,g_J(X_f,X_h).
 \label{eq:covmetric}
\end{equation}
In particular $(\Delta f)^2_J=\frac\hbar2\|X_f\|_{g_J}^2$: the same statement as on the space of states, now on the classical phase space. A relation of this type underlies the geometric discussion of cosmological squeezing in Ref.~\cite{ack}. Writing $f=l^Tx$ and $X_f=\Pi l$, the covariance matrix of the coordinates is
\begin{equation}
 V_J=\frac{\hbar}{2}\Pi^Tg_J\Pi=\frac{\hbar}{2}g_J^{-1},
 \label{eq:dual}
\end{equation}
where the second equality follows from Eq.~\eqref{eq:matrixmetric}. Coordinate fluctuations are governed by the \emph{inverse} metric because observables are covectors, mapped to vectors by the symplectic form before the metric is applied. A large coefficient of $\dd q^2$ thus means a \emph{small} $q$ variance, since $X_q$ points along the momentum direction. The complete two-point function of centered linear observables is
\begin{equation}
 \langle\hat f\hat h\rangle_J
 =\frac{\hbar}{2}g_J(X_f,X_h)+\frac{i\hbar}{2}\{f,h\}.
 \label{eq:twopoint}
\end{equation}
The symplectic part fixes the commutator and the metric part specifies the state.

\subsection{The bridge: coherent states}

The parallel between $(\mathcal P,\Om_{\mathcal H},G)$ and $(\Gamma,\Om,g_J)$ is more than an analogy. For a phase-space displacement $\gamma$ define the Weyl operator
\begin{equation}
 W(\gamma)=\exp\!\left[\frac{i}{\hbar}\Om(\gamma,\hat x)\right],
 \label{eq:weyl}
\end{equation}
and the coherent states $|\gamma;J\rangle=W(\gamma)\vac{J}$ \cite{perelomov}. The map $\gamma\mapsto|\gamma;J\rangle$ embeds the classical phase space in the space of quantum states.

\begin{proposition}
\label{prop:coherent}
On the coherent states of $\vac{J}$, the quantum geometric tensor $h=\langle\dd\psi|\dd\psi\rangle-\langle\dd\psi|\psi\rangle\langle\psi|\dd\psi\rangle$ is
\begin{equation}
 h(\dd\gamma_1,\dd\gamma_2)=\frac{1}{2\hbar}\bigl[g_J(\dd\gamma_1,\dd\gamma_2)-i\,\Om(\dd\gamma_1,\dd\gamma_2)\bigr].
 \label{eq:coherenth}
\end{equation}
In particular the Fubini--Study metric restricted to coherent states is $g_J/2\hbar$, and its imaginary part, the Berry curvature, is $\Om/2\hbar$.
\end{proposition}

\begin{proof}
Up to a phase, removed by the projection in $h$, a displacement changes the state by $\dd\psi=(i/\hbar)\,\Om(\dd\gamma,\hat x-\gamma)\psi$. The operator $\Om(\dd\gamma,\hat x-\gamma)$ is a centered linear observable with Hamiltonian vector field $-\dd\gamma$, so Eq.~\eqref{eq:twopoint} gives $h(\dd\gamma_1,\dd\gamma_2)=\hbar^{-2}\bigl[\frac\hbar2g_J(\dd\gamma_1,\dd\gamma_2)+\frac{i\hbar}2\{\Om(\dd\gamma_1,\cdot),\Om(\dd\gamma_2,\cdot)\}\bigr]$, and $\{\Om(a,\cdot),\Om(b,\cdot)\}=-\Om(a,b)$.
\end{proof}

This is the coherent-state realization of the K\"ahler structure of quantum mechanics \cite{kibble,provost,ashsch}. Read as a statement about vacua, it says that choosing a vacuum is choosing which quantum metric the classical phase space inherits. Consistently with Eq.~\eqref{eq:AAspeed}, the linear observable $f$ moves coherent states along $X_f$ at Fubini--Study speed $\|X_f\|_{g_J}/\sqrt{2\hbar}=\Delta f/\hbar$. Section~\ref{sec:rays} completes the picture: varying $g$ itself moves the vacuum in directions orthogonal to the displacements, with a metric of its own.

\subsection{Three descriptions of a Gaussian vacuum}
\label{sec:dictionary}

Table~\ref{tab:dictionary} collects the covariance description standard in quantum optics \cite{smd,arvind}, the complex-structure description of Refs.~\cite{am,bhy,hb}, in which the relative complex structure is $\Delta=-JJ_0$, and the metric description developed here. Entries referring to a reference vacuum $J_0$, with metric $g_0$ and covariance $V_0$, are derived in Section~\ref{sec:squeezed}.

\begin{table}[ht]
\centering
\small
\renewcommand{\arraystretch}{1.35}\setlength{\tabcolsep}{4pt}
\begin{tabular}{@{}>{\raggedright\arraybackslash}p{0.15\textwidth}>{\raggedright\arraybackslash}p{0.28\textwidth}>{\raggedright\arraybackslash}p{0.22\textwidth}>{\raggedright\arraybackslash}p{0.28\textwidth}@{}}
\toprule
 & Covariance $V$ & Complex structure $J$ & Metric $g$\\
\midrule
State & $V>0$, symplectic eigenvalues $\hbar/2$ & $J^2=-\Id$, $J^T\Om J=\Om$, $\Om(J\cdot,\cdot)>0$ & $g>0$, $g\Pi g=-\Om$\\
Relations & $V=\frac{\hbar}{2}g^{-1}$ & $J=-\Pi g$ & $g=-\Om J$\\
Variance of $f$ & $l^TVl$, \ $f=l^Tx$ & $\frac{\hbar}{2}\Om(JX_f,X_f)$ & $\frac{\hbar}{2}\|X_f\|_g^2$\\
Comparison with reference & $V_0V^{-1}$ & $\Delta=-JJ_0=L^{-1}$ & $L=g_0^{-1}g$\\
Squeezing parameters & $\spec(V_0V^{-1})$ & $\spec\Delta$ & $\spec L=\{e^{\pm2r_i}\}$\\
Reference occupation & $\frac14\bigl[\Tr(V_0^{-1}V)-2N\bigr]$ & $\frac14(\Tr\Delta-2N)$ & $\tfrac14(\Tr L-2N)=\sum_i\sinh^2r_i$\\
Evolution by $S$ & $SVS^T$ & $SJS^{-1}$ & $S^{-T}gS^{-1}$\\
Overlap with reference & \mbox{$\det\!\bigl[\tfrac{\hbar}{4}(V^{-1}+V_0^{-1})\bigr]^{-1/2}$} & --- & \mbox{$\det\!\bigl[\tfrac12(g+g_0)\bigr]^{-1/2}$}\\
\bottomrule
\end{tabular}
\caption{Three equivalent descriptions of a centered pure Gaussian state, in physical Darboux coordinates where $\det g=1$. The overlap is the squared modulus $|\langle0;g_0|0;g\rangle|^2$; see Eq.~\eqref{eq:overlap}.}
\label{tab:dictionary}
\end{table}

\section{The harmonic oscillator}
\label{sec:oscillator}

\subsection{Complex structures from mode functions}
\label{sec:modes}

For one canonical pair, take the oscillator $H_0=p^2/(2m)+m\omega^2q^2/2$. Its phase space is in one-to-one correspondence with the space of solutions of $\ddot x=-\omega^2x$, through the initial data $(q,p)=(x,m\dot x)|_{t_0}$. A complex structure on the space of solutions is specified by a complex mode function $f$, with $x=af+\bar a\bar f$, normalized by
\begin{equation}
 -im\bigl(f\dot{\bar f}-\dot f\bar f\bigr)=\hbar .
 \label{eq:wronsk}
\end{equation}
The induced complex structure $J$ on $\Gamma$ has the initial data of $f$ as its $+i$ eigenvector, and the corresponding annihilation variable is the linear function
\begin{equation}
 a=\frac1\hbar\,\Om\bigl(J(\bar f,m\dot{\bar f})_{t_0},(q,p)\bigr)=-\frac i\hbar\bigl(m\dot{\bar f}\,q-\bar f\,p\bigr)\Big|_{t_0}.
 \label{eq:annihilation}
\end{equation}
Every compatible complex structure has the form
\begin{equation}
 J=\begin{pmatrix}A&B\\D&-A\end{pmatrix},\qquad A^2+BD=-1,\quad B<0,\quad D>0,
\end{equation}
where $J^2=-\Id$ forces the trace-free form and $A^2+BD=-1$, and positivity of $g_J$ forces the signs of $B$ and $D$. In terms of the mode function,
\begin{equation}
 A=\frac m\hbar\bigl(f\dot{\bar f}+\dot f\bar f\bigr),\qquad
 B=-\frac2\hbar\,f\bar f,\qquad
 D=\frac{2m^2}\hbar\,\dot f\dot{\bar f},
 \label{eq:ABD}
\end{equation}
evaluated at $t_0$; the condition $A^2+BD=-1$ is the normalization~\eqref{eq:wronsk}. The metric and covariance are
\begin{equation}
 g_J=\begin{pmatrix}D&-A\\-A&-B\end{pmatrix},
 \qquad V_J=\frac{\hbar}{2}\begin{pmatrix}-B&A\\A&D\end{pmatrix},
 \label{eq:oscmetric}
\end{equation}
so that, using $g_J(X_q,X_q)=-B$ and $g_J(X_p,X_p)=D$,
\begin{equation}
 (\Delta q)^2=-\frac{\hbar B}{2}=|f|^2,\qquad (\Delta p)^2=\frac{\hbar D}{2}=m^2|\dot f|^2,\qquad C_{qp}=\frac{\hbar A}{2}=m\operatorname{Re}(f\dot{\bar f}),
 \label{eq:oscfluct}
\end{equation}
and
\begin{equation}
 (\Delta q)^2(\Delta p)^2-C_{qp}^2=\frac{\hbar^2}{4},
 \qquad\text{equivalently}\qquad (\Delta q)^2(\Delta p)^2=\frac{\hbar^2}{4}(1+A^2).
 \label{eq:robertson}
\end{equation}
In these physical Darboux coordinates $\det g_J=1$ and $\det V_J=\hbar^2/4$.

With $\alpha=(m\omega/\hbar)^{1/2}$, the dimensionless variables $Q=\alpha q$ and $P=p/(\hbar\alpha)$ give $[\hat Q,\hat P]=i$ and $\Om=\hbar\,\dd P\wedge\dd Q$. The positive-frequency solution $f=e^{-i\omega t}/(\alpha\sqrt2)$ gives, at $t_0=0$, the annihilation variable $a=(Q+iP)/\sqrt2$ and
\begin{equation}
 J_0=\begin{pmatrix}0&-1\\1&0\end{pmatrix},
 \qquad g_0=\hbar(\dd Q^2+\dd P^2),\qquad V_0=\tfrac12\Id
 \label{eq:standard}
\end{equation}
in the variables $(Q,P)$. This is the usual vacuum: its fluctuations are evenly distributed between $Q$ and $P$ because the metric is isotropic, and the curve $g_0(X,X)=\hbar$ is a circle. Writing $g_J=\hbar K_J$ in these coordinates gives $V_J=K_J^{-1}/2$, with $\det K_J=1$. Isotropy is a statement relative to the oscillator scaling: every compatible metric is Euclidean in suitable canonical coordinates. ``Circular'' and ``elliptical'' are therefore not, by themselves, invariant characterizations of a state; Theorem~\ref{thm:classification} below makes this precise.

\subsection{The vacuum in the Schr\"odinger representation}

Since $\hat a\Psi_0=0$, Eq.~\eqref{eq:annihilation} together with $f\bar f=-\frac\hbar2B$ and $mf\dot{\bar f}=\frac\hbar2(A+i)$, which follow from Eqs.~\eqref{eq:wronsk} and~\eqref{eq:ABD}, gives
\begin{equation}
 \hat p\,\Psi_0=-B^{-1}(A+i)\,\hat q\,\Psi_0 .
 \label{eq:pvac}
\end{equation}
In the representation $\hat q\Psi=q\Psi$, $\hat p\Psi=(-i\hbar\partial_q+\xi q)\Psi$, the solution is
\begin{equation}
 \Psi_0(q)=\exp\!\left\{\frac1{2\hbar}\bigl[B^{-1}(1-iA)-i\xi\bigr]q^2\right\}.
 \label{eq:xivac}
\end{equation}
For $\xi=0$ this is the normalized Gaussian
\begin{equation}
 \psi_J(q)=\bigl[\pi(-\hbar B)\bigr]^{-1/4}\exp\!\left[\frac{1-iA}{2\hbar B}q^2\right],
 \label{eq:wavefunction}
\end{equation}
whose modulus fixes the $q$ variance and whose quadratic phase fixes the $q$--$p$ correlation; this is the role of the parameter $A$. For $\xi=-B^{-1}(A+i)$ the vacuum is the constant function $\Psi_0=1$. That representation is unitarily equivalent to the first once the Gaussian factor is transferred into the measure: with $\dd\mu_J(q)=\exp[q^2/(\hbar B)]\,\dd q/\sqrt{\pi(-\hbar B)}$ and $(U\phi)(q)=\psi_J(q)\phi(q)$, the map $U:L^2(\dd\mu_J)\to L^2(\dd q)$ is unitary, $U1=\psi_J$, and
\begin{equation}
 U^{-1}\hat pU=-i\hbar\partial_q-B^{-1}(A+i)q .
\end{equation}
The complex multiplicative term is compatible with self-adjointness because the operator acts on $L^2(\dd\mu_J)$ rather than on $L^2(\dd q)$. The field-theoretic counterpart, relating Fock and Gaussian-measure Schr\"odinger representations, is treated in Refs.~\cite{ccq,ccq04}.

\subsection{Weyl operators and a positive phase-space distribution}

The linear observable $\Om(\gamma,x)$ has Hamiltonian vector field $-\gamma$, so by Eq.~\eqref{eq:covmetric} its variance is $\frac\hbar2g_J(\gamma,\gamma)$, and the vacuum expectation values of the Weyl operators~\eqref{eq:weyl} are
\begin{equation}
 \langle W(\gamma)\rangle_J=\exp\!\left[-\frac{g_J(\gamma,\gamma)}{4\hbar}\right].
 \label{eq:weylstate}
\end{equation}
They are constant on the level curves of $g_J$: circles for $J_0$, ``squeezed circles'' for other complex structures. Equation~\eqref{eq:weylstate} defines a positive state on the Weyl algebra, whose GNS representation is the vacuum representation of $J$. Its Fourier transform is the Wigner density, which in physical Darboux coordinates is
\begin{equation}
 \rho_J(x)=\frac{1}{(\pi\hbar)^N}\exp\!\left[-\frac{g_J(x,x)}{\hbar}\right],
 \qquad \int\rho_J(x)\,\dd^Nq\,\dd^Np=1,
 \label{eq:wigner}
\end{equation}
using $\det g_J=1$; in the variables of Eq.~\eqref{eq:standard}, $\rho_0=\pi^{-1}e^{-(Q^2+P^2)}$. The exponent is the metric norm of the phase-space point. This is a positive probability density on the classical phase space whose moments are exactly the Weyl-ordered vacuum expectation values, $\langle(\hat x^n\hat p^m)_W\rangle_J=\langle x^np^m\rangle_{\rho_J}$ \cite{ack}, and its covariance is Eq.~\eqref{eq:dual}. The metric thus determines the uncertainties, the Weyl characteristic function, the vacuum selected by the algebraic state, and a classical distribution reproducing its Weyl-ordered moments. That distribution does not, however, reproduce operator products in other orderings, nor does it imply collapse, decoherence or a selected outcome; its scope is analyzed in Ref.~\cite{corichicl}.

\subsection{A puzzle about rotations}
\label{sec:rotation}

Let $\Gamma$ carry the standard $J_0$ and apply $T=R(\theta)S(\lambda)$, with $R(\theta)$ the rotation by $\theta$ in the $(Q,P)$ plane and $S(\lambda)=\diag(\lambda,\lambda^{-1})$. The induced complex structure $J=TJ_0T^{-1}$ is, with $c=\cos\theta$ and $s=\sin\theta$,
\begin{equation}
 J=\begin{pmatrix}
 (\lambda^2-\lambda^{-2})cs&-(\lambda^2c^2+\lambda^{-2}s^2)\\
 \lambda^{-2}c^2+\lambda^2s^2&-(\lambda^2-\lambda^{-2})cs
 \end{pmatrix},
\end{equation}
and the fluctuations are
\begin{equation}
 (\Delta Q)^2=\tfrac12(\lambda^2c^2+\lambda^{-2}s^2),\quad
 (\Delta P)^2=\tfrac12(\lambda^{-2}c^2+\lambda^2s^2),\quad
 C_{QP}=\tfrac12(\lambda^2-\lambda^{-2})cs.
\end{equation}
For $\lambda\neq1$ the parameter $A$ of Eq.~\eqref{eq:ABD} is nonzero unless $\theta$ is a multiple of $\pi/2$, and then
\begin{equation}
 (\Delta Q)^2(\Delta P)^2=\frac14+C_{QP}^2=\frac14\Bigl[1+(\lambda^2-\lambda^{-2})^2c^2s^2\Bigr]>\frac14 .
\end{equation}
A rotation of a minimum-uncertainty state seems to produce a state that is no longer of minimum uncertainty. Since $R$ commutes with $J_0$, it is not even clear why the order of $R$ and $S$ matters.

The resolution is that the product of the displayed variances is not the right measure. The rotated state is the same squeezed vacuum seen along tilted axes, and the rotation introduces a covariance $C_{QP}$ between the displayed observables. The full Robertson--Schr\"odinger relation \cite{schrodinger,robertson}, Eq.~\eqref{eq:robertson}, involves the determinant of the covariance, which stays fixed:
\begin{equation}
 (\Delta Q)^2(\Delta P)^2-C_{QP}^2=\det V=\frac14 .
\end{equation}
Every centered pure Gaussian vacuum saturates it; only the subfamily $A=0$ also saturates the simpler bound $\Delta Q\,\Delta P\ge\frac12$ in the given axes. Geometrically, the product $\Delta Q\,\Delta P$ measures the box circumscribing the uncertainty ellipse, while the area of the ellipse is fixed by purity. As for the order: $SR$ gives the same state as $S$, because $R$ preserves $J_0$, while $RS$ rotates an already squeezed ellipse. For $\lambda=2$ and $\theta=\pi/4$, both displayed variances equal $17/16$, $C_{QP}=15/16$, and the principal variances are $2$ and $1/8$ (Figure~\ref{fig:ellipses}, right). Equal variances along $Q$ and $P$ coexist with strong squeezing along rotated quadratures, and reading only the diagonal of the covariance matrix can hide squeezing entirely. Section~\ref{sec:rayleigh} gives the invariant statement.

\section{Squeezed vacua}
\label{sec:squeezed}

\subsection{Two things one can do with a canonical transformation}
\label{sec:operations}

Squeezed vacua arise from a symplectic transformation of $\Gamma$. Take for simplicity $Q'=\alpha Q$, $P'=\alpha^{-1}P$, which elongates one axis and squeezes the other while preserving area. With it one can do two quite different things.

\emph{(i) Rewrite the metric in the new coordinates.} The standard metric becomes $g_0=\hbar(\alpha^{-2}\dd Q'^2+\alpha^2\dd P'^2)$. This is the same tensor and the same vacuum in another description; if observables are relabelled consistently, no prediction changes. In general, for $x'=Tx$, $J'_0=TJ_0T^{-1}$ and $g'_0=T^{-T}g_0T^{-1}$.

\emph{(ii) Declare a new metric that looks Euclidean in the new coordinates.} Define
\begin{equation}
 g'=\hbar(\dd Q'^2+\dd P'^2)=\hbar(\alpha^2\dd Q^2+\alpha^{-2}\dd P^2),
\end{equation}
and in general, for $x'=Tx$,
\begin{equation}
 g_{\rm new}=T^Tg_0T,\qquad J_{\rm new}=T^{-1}J_0T .
 \label{eq:reset}
\end{equation}
Since a metric defines a quantum theory and a vacuum, this selects a new vacuum $\vac{J'}$, different from the original one when $\alpha\neq1$. An ``observer'' who uses $(Q',P')$ will regard $g'$ and its vacuum, $\Psi'_0\propto e^{-Q'^2/2}$, as the natural ones. The original observer sees the curve $g'(X,X)=\hbar$ as an ellipse, a squeezed circle, and computes from Eq.~\eqref{eq:covmetric}
\begin{equation}
 (\Delta Q)^2_{g'}=\frac{1}{2\alpha^2},\qquad (\Delta P)^2_{g'}=\frac{\alpha^2}{2}.
 \label{eq:slidesvac}
\end{equation}
For $\alpha>1$ the observable $Q$ is squeezed and $P$ is stretched, with the product of uncertainties unchanged.

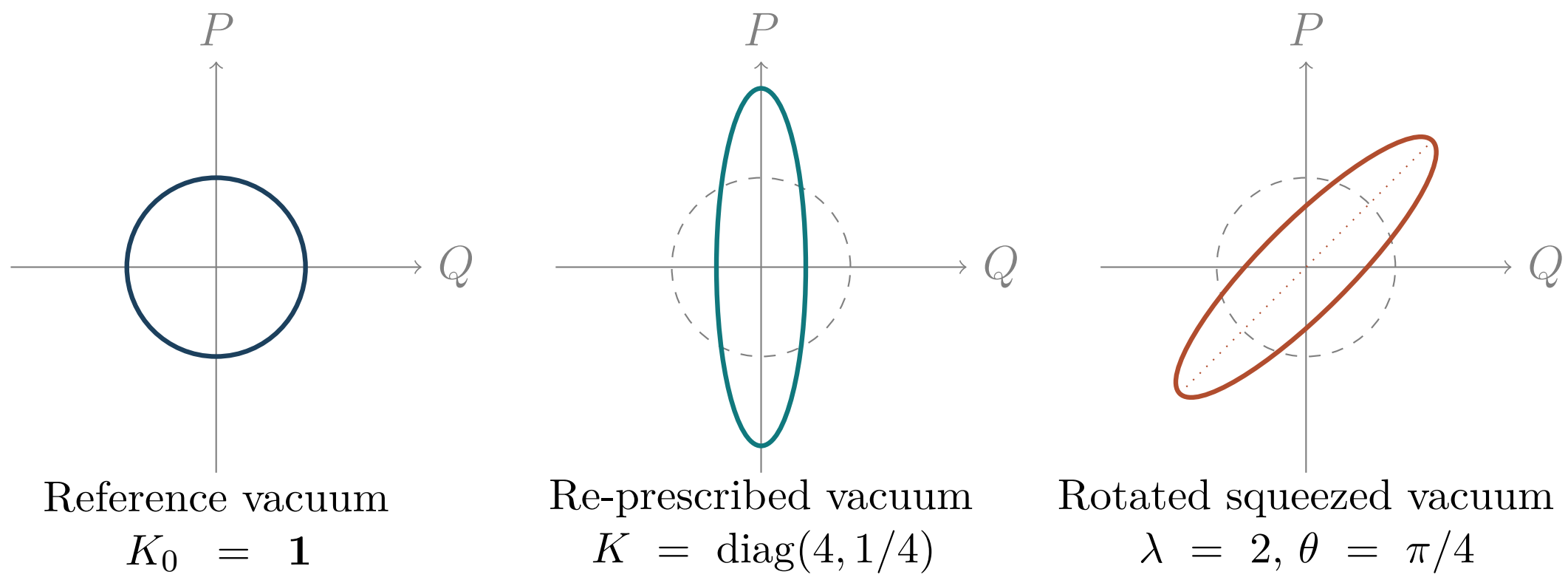
\begin{figure}[t]
\centering
\begin{tikzpicture}[scale=0.8]
\begin{scope}[xshift=0cm]
 \draw[->,gray] (-2.3,0)--(2.3,0) node[right]{$Q$};
 \draw[->,gray] (0,-2.3)--(0,2.3) node[above]{$P$};
 \draw[deepblue,very thick] (0,0) circle (1);
 \node[align=center,text width=4.8cm,font=\small] at (0,-2.9) {Reference vacuum\\$K_0=\Id$};
\end{scope}
\begin{scope}[xshift=6.1cm]
 \draw[->,gray] (-2.3,0)--(2.3,0) node[right]{$Q$};
 \draw[->,gray] (0,-2.3)--(0,2.3) node[above]{$P$};
 \draw[gray,dashed] (0,0) circle (1);
 \draw[teal,very thick] (0,0) ellipse (0.5 and 2);
 \node[align=center,text width=4.8cm,font=\small] at (0,-2.9) {Re-prescribed vacuum\\$K=\diag(4,1/4)$};
\end{scope}
\begin{scope}[xshift=12.2cm]
 \draw[->,gray] (-2.3,0)--(2.3,0) node[right]{$Q$};
 \draw[->,gray] (0,-2.3)--(0,2.3) node[above]{$P$};
 \draw[gray,dashed] (0,0) circle (1);
 \draw[rust,very thick,rotate=45] (0,0) ellipse (2 and 0.5);
 \draw[rust,dotted,rotate=45] (-2,0)--(2,0);
 \node[align=center,text width=4.8cm,font=\small] at (0,-2.9) {Rotated squeezed vacuum\\$\lambda=2$, $\theta=\pi/4$};
\end{scope}
\end{tikzpicture}
\caption{Metric contours $g_J(x,x)=\hbar$, equivalently $x^TV_J^{-1}x=2$, in a common reference plane. Each ellipse has area $\pi$ in dimensionless coordinates; the reference circle is dashed in the latter panels. These are Gaussian contours, not quantum trajectories.}
\label{fig:ellipses}
\end{figure}

A third operation, (iii) \emph{active squeezing}, pushes the state forward by $S=\diag(\lambda,\lambda^{-1})$ with fixed observables:
\begin{equation}
 J_{\rm act}=SJ_0S^{-1},\qquad
 g_{\rm act}=S^{-T}g_0S^{-1},\qquad
 V_{\rm act}=SV_0S^T,
 \label{eq:active}
\end{equation}
with $(\Delta Q)^2=\lambda^2/2$ and $(\Delta P)^2=\lambda^{-2}/2$. Re-prescription by $T$ and active squeezing by $T^{-1}$ produce the same pair of metrics, which is why Eqs.~\eqref{eq:slidesvac} and~\eqref{eq:active} squeeze opposite quadratures when $\alpha=\lambda$. Figure~\ref{fig:ellipses} shows the metric contours, and Figure~\ref{fig:disk} below the three operations in the space of vacua.

\subsection{Relative vacua}

The discussion shows that ``squeezed'' always means ``squeezed with respect to a reference''.

\begin{definition}
A \emph{relative vacuum} is a pair $(g_0,g)$ of compatible metrics on $(\Gamma,\Om)$, the first regarded as the reference and the second as the state. Two relative vacua $(g_0,g)$ and $(g_0',g')$ are \emph{equivalent} if a symplectic map $T$ satisfies $g_0'=T^{-T}g_0T^{-1}$ and $g'=T^{-T}gT^{-1}$.
\end{definition}

Equivalent pairs differ only by a common re-description, operation (i). The comparison operator
\begin{equation}
 L:=g_0^{-1}g=-J_0J
 \label{eq:L}
\end{equation}
is self-adjoint and positive with respect to $g_0$, since $g_0(u,Lv)=g(u,v)$. Substituting $L=-J_0J$ and using $J_0^2=J^2=-\Id$ gives
\begin{equation}
 J_0 L J_0^{-1}=L^{-1},
 \qquad
 \spec(L)=\{e^{\pm2r_1},\ldots,e^{\pm2r_N}\},\quad r_i\ge0,
 \label{eq:reciprocal}
\end{equation}
which defines the relative squeezing parameters $r_i$. The relative complex structure $\Delta=-JJ_0$ of Ref.~\cite{hb} equals $L^{-1}$. If $Lu=e^{2r}u$, then $L(J_0u)=e^{-2r}J_0u$ and $\Om(u,J_0u)=-g_0(u,u)\neq0$, so the two directions can be normalized to a canonical pair: reduced and enlarged uncertainties are complementary as a consequence of compatibility, not as an extra assumption about an ellipse.

\subsection{Classification}

\begin{theorem}[normal form of relative vacua]
\label{thm:classification}
Two relative vacua are equivalent if and only if their relative metrics have the same spectrum. Every relative vacuum is equivalent to
\begin{equation}
 g_0=\hbar\sum_{i=1}^N(\dd Q_i^2+\dd P_i^2),\qquad
 g=\hbar\sum_{i=1}^N\bigl(e^{2r_i}\dd Q_i^2+e^{-2r_i}\dd P_i^2\bigr),
 \label{eq:normalform}
\end{equation}
with $r_1\ge\cdots\ge r_N\ge0$ and $\Om=\hbar\sum_i\dd P_i\wedge\dd Q_i$. The equivalence classes form the double coset $U(N)\backslash\Sp(2N,\R)/U(N)$.
\end{theorem}

This is the Bloch--Messiah (Cartan) decomposition \cite{arvind,braunstein} restated for pairs of metrics; we include a short proof in our notation.

\begin{proof}
Under $x'=Tx$ both metrics transform as $g\mapsto T^{-T}gT^{-1}$, so $L\mapsto TLT^{-1}$ and the spectrum is invariant. Conversely, choose canonical coordinates in which $g_0$ is Euclidean. Then $K=g/\hbar$ is the matrix of $L$; it is symmetric, positive and symplectic by Eq.~\eqref{eq:matrixmetric}, so $K=\exp X$ with $X$ symmetric and $XJ_0+J_0X=0$. The stabilizer of $g_0$ is $U(N)=\Sp(2N,\R)\cap O(2N)$, and an element of it brings $X$ to $\diag(2r_1,\ldots,2r_N,-2r_1,\ldots,-2r_N)$ \cite{arvind}. Permutations of canonical pairs and the maps $(Q_i,P_i)\mapsto(P_i,-Q_i)$ lie in $U(N)$ and order the $r_i\ge0$. The normal form depends only on $\spec L$. Finally, compatible metrics form $\Sp(2N,\R)/U(N)$ \cite{siegel,bhy}, and orbits of the diagonal action on pairs are orbits of the stabilizer of $g_0$ on the second factor.
\end{proof}

A single compatible metric has no invariants, since each is Euclidean in suitable canonical coordinates; a pair has exactly $N$. Any coordinate-independent comparison of two vacua is therefore a symmetric function of the $r_i$.

\begin{corollary}
\label{cor:invariants}
For a relative vacuum with squeezing parameters $r_i$:
\begin{enumerate}
\item the reference occupation number is $\langle\hat N_0\rangle_J=\tfrac14(\Tr L-2N)=\tfrac18\Tr(L+L^{-1}-2)=\sum_i\sinh^2r_i$;
\item the squared overlap is $|\langle0;g_0|0;g\rangle|^2=\prod_i(\cosh r_i)^{-1}$;
\item the extremal variance ratios of linear observables are $e^{\mp2r_{\max}}$;
\item the Fubini--Study length of the shortest path from $g_0$ to $g$ within the Gaussian family is $\bigl(\tfrac12\sum_ir_i^2\bigr)^{1/2}$.
\end{enumerate}
\end{corollary}

\begin{proof}
Each quantity is invariant under common symplectic maps, so it may be computed in the normal form. There $\hat N_0=\sum_i(\hat Q_i^2+\hat P_i^2-1)/2$ and $V=K^{-1}/2$, which gives item~1 since $\Tr L=\Tr L^{-1}$. Items~2, 3 and~4 follow from Eq.~\eqref{eq:overlap}, Section~\ref{sec:rayleigh} and Section~\ref{sec:distances}.
\end{proof}

\begin{figure}[t]
\centering
\begin{tikzpicture}[scale=2.0]
\begin{scope}[shift={(0.0,0)}]
\fill[gray!6] (0,0) circle (1); \draw[gray] (0,0) circle (1);
\draw[gray!70,dashed] (0.000,0.000) -- (0.550,0.000);
\fill[gray!60] (0.000,0.000) circle (0.03); \fill[gray!60] (0.550,0.000) circle (0.03);
\draw[->,deepblue!70,thin,shorten >=2.5pt] (0.000,0.000) -- (-0.461,0.004);
\draw[->,rust!70,thin,shorten >=2.5pt] (0.550,0.000) -- (-0.135,0.416);
\draw[teal,very thick] plot[smooth] coordinates {(-0.461,0.004) (-0.433,0.028) (-0.405,0.054) (-0.377,0.081) (-0.349,0.110) (-0.321,0.141) (-0.293,0.174) (-0.265,0.209) (-0.238,0.246) (-0.211,0.285) (-0.185,0.326) (-0.159,0.370) (-0.135,0.416)};
\fill[deepblue] (-0.461,0.004) circle (0.045); \fill[rust] (-0.135,0.416) circle (0.045);
\node[font=\small,align=center] at (0,-1.22) {(a) passive re-description\\[1pt]$r=0.62$};
\end{scope}
\begin{scope}[shift={(2.55,0)}]
\fill[gray!6] (0,0) circle (1); \draw[gray] (0,0) circle (1);
\draw[gray!70,dashed] (0.000,0.000) -- (0.550,0.000);
\fill[gray!60] (0.000,0.000) circle (0.03); \fill[gray!60] (0.550,0.000) circle (0.03);
\draw[->,deepblue!70,thin,shorten >=2.5pt] (0.000,0.000) -- (-0.187,-0.409);
\draw[teal,very thick] plot[smooth] coordinates {(-0.187,-0.409) (-0.148,-0.376) (-0.106,-0.343) (-0.061,-0.309) (-0.012,-0.274) (0.041,-0.239) (0.098,-0.204) (0.159,-0.168) (0.226,-0.133) (0.297,-0.098) (0.375,-0.064) (0.459,-0.031) (0.550,-0.000)};
\fill[deepblue] (-0.187,-0.409) circle (0.045); \fill[rust] (0.550,0.000) circle (0.045);
\node[font=\small,align=center] at (0,-1.22) {(b) re-prescription\\[1pt]$r=0.97$};
\end{scope}
\begin{scope}[shift={(5.1,0)}]
\fill[gray!6] (0,0) circle (1); \draw[gray] (0,0) circle (1);
\draw[gray!70,dashed] (0.000,0.000) -- (0.550,0.000);
\fill[gray!60] (0.000,0.000) circle (0.03); \fill[gray!60] (0.550,0.000) circle (0.03);
\draw[->,rust!70,thin,shorten >=2.5pt] (0.550,0.000) -- (0.485,0.611);
\draw[teal,very thick] plot[smooth] coordinates {(0.000,0.000) (0.040,0.051) (0.081,0.102) (0.121,0.153) (0.162,0.204) (0.202,0.255) (0.242,0.305) (0.283,0.356) (0.323,0.407) (0.364,0.458) (0.404,0.509) (0.444,0.560) (0.485,0.611)};
\fill[deepblue] (0.000,0.000) circle (0.045); \fill[rust] (0.485,0.611) circle (0.045);
\node[font=\small,align=center] at (0,-1.22) {(c) active squeezing\\[1pt]$r=1.05$};
\end{scope}
\end{tikzpicture}
\caption{Single-mode vacua as points of the disk $z=e^{2i\theta}\tanh r$ with metric $\dd s^2=|\dd z|^2/[2(1-|z|^2)^2]$, Eq.~\eqref{eq:FSsqueeze}. Gray: an initial relative vacuum, with reference at the centre and state on the real axis. Blue: reference; red: state; teal: the geodesic joining them. (a) Operation (i), a passive re-description, is an isometry and moves both points, preserving the squeezing $r$. (b) Operation (ii), re-prescription, moves only the reference. (c) Operation (iii), active squeezing, moves only the state. The value of $r$ printed below each panel is the only invariant of the pair (Theorem~\ref{thm:classification}).}
\label{fig:disk}
\end{figure}
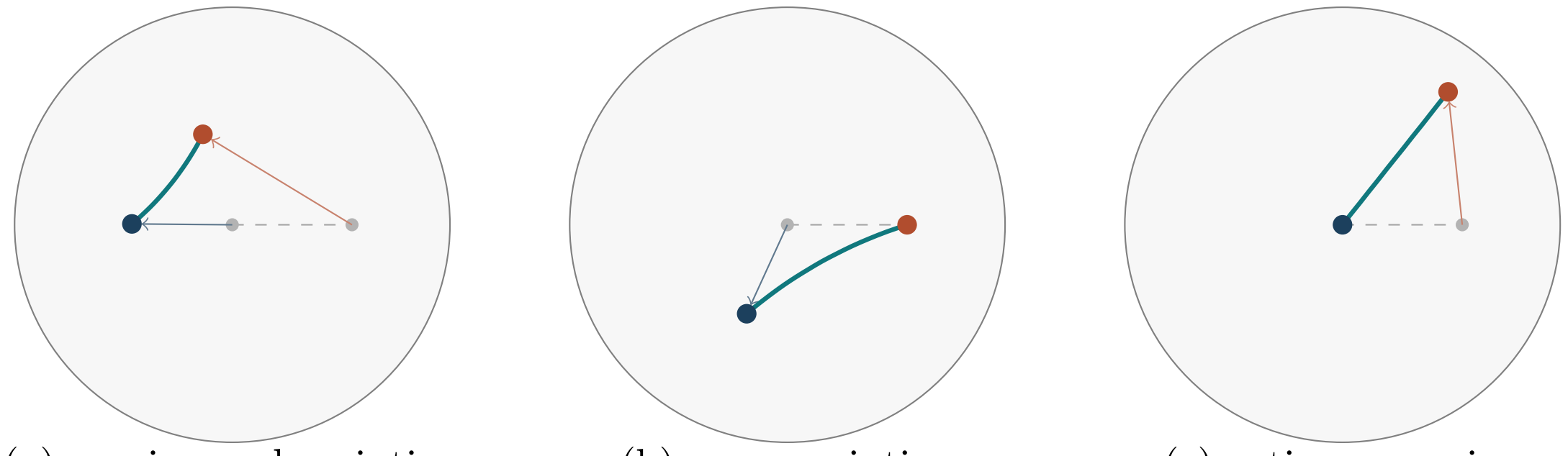

For a single pair the invariant $r$ is, up to a factor $1/\sqrt2$, the hyperbolic distance between reference and state in the space of vacua, which is a disk (Section~\ref{sec:rays}). Figure~\ref{fig:disk} shows the three operations of Section~\ref{sec:operations} in this picture. For the rotated oscillator of Section~\ref{sec:rotation}, $\Tr L=\lambda^2+\lambda^{-2}$ and $r=|\log\lambda|$ independently of $\theta$: rotation moves the state around the reference at fixed distance, which is the invariant answer to the puzzle.

\subsection{One observable, two metrics}
\label{sec:rayleigh}

The relative spectrum describes the whole state. For a specified nonzero linear observable $f$, the variance ratio is the Rayleigh quotient of $L$,
\begin{equation}
 \mathcal Q_f:=\frac{(\Delta f)^2_J}{(\Delta f)^2_0}
 =\frac{g(X_f,X_f)}{g_0(X_f,X_f)}
 =\frac{g_0(X_f,LX_f)}{g_0(X_f,X_f)},
 \qquad e^{-2r_{\max}}\le\mathcal Q_f\le e^{2r_{\max}},
 \label{eq:rayleigh}
\end{equation}
with both bounds attained, since every phase-space direction is the Hamiltonian vector field of a linear observable. For $J\neq J_0$ some observable is squeezed and another anti-squeezed; whether a preselected laboratory observable is squeezed is answered by inserting its Hamiltonian vector field in Eq.~\eqref{eq:rayleigh}. The rotated oscillator of Section~\ref{sec:rotation} is squeezed relative to the reference although neither displayed variance shows it.

The Rayleigh quotient compares one observable in two metrics. Its counterpart within a single metric is the angle $\theta_{fh}$ between $X_f$ and $X_h$ in $g$. By Eq.~\eqref{eq:covmetric}, $\cos\theta_{fh}$ is the correlation coefficient of $f$ and $h$, and Ref.~\cite{corichicl} shows that for a canonical pair $|\sin\theta_{fh}|$ is the size of the commutator relative to the fluctuations, which controls the departure of arbitrarily ordered correlation functions from the moments of $\rho_J$. Lengths in two metrics measure squeezing; angles in one metric measure correlation and non-commutativity.

\subsection{The vacuum of another observer}
\label{sec:observer}

For the unrotated active oscillator, $\lambda=e^r$ and
\begin{equation}
 \hat a_J=\frac{e^{-r}\hat Q+i e^r\hat P}{\sqrt2}
 =\cosh r\,\hat a_0-\sinh r\,\hat a_0^\dagger,
 \qquad \hat a_J\vac{J}=0.
\end{equation}
The state has no $J$-particles and $\sinh^2r$ reference particles: ``vacuum'' and ``squeezed'' are compatible descriptions once the reference is stated. Every compatible $J$ defines the positive quadratic Hamiltonian $H_J(x)=\frac{\omega}{2}g_J(x,x)$, whose symmetrically ordered quantization has $\vac{J}$ as ground state, since in $J$-adapted canonical coordinates $\hat H_J=\hbar\omega\sum_i(\hat a_{J,i}^\dagger\hat a_{J,i}+1/2)$. This does not make $\vac{J}$ the ground state of an independently fixed Hamiltonian. The ``observer'' of operation (ii) is therefore, so far, a vacuum prescription; turning it into an observer requires tying it to a time generator, a mode selection or a measurement protocol (Section~\ref{sec:discussion}).

\section{Evolution as a deformation of the metric}
\label{sec:dynamics}

Let $S(t,t_0)$ be the classical propagator of a quadratic, possibly time-dependent Hamiltonian, and $F=\dot SS^{-1}$ its generator. In the active description
\begin{equation}
 g_t=S^{-T}g_{t_0}S^{-1},\qquad
 \dot J=[F,J],\qquad \dot g=-F^Tg-gF,
 \qquad \dot V=FV+VF^T.
 \label{eq:evolution}
\end{equation}
There is an equivalent covariant, or Heisenberg, description, which is the one used in Ref.~\cite{ack}. Let $I_t$ evaluate each classical solution at time $t$. A fixed metric $g_{\rm sol}$ on the space of solutions induces $g_t(u,v)=g_{\rm sol}(I_t^{-1}u,I_t^{-1}v)$; equivalently, the metric stays fixed and the time-dependent observables $f_t=f\circ I_t$ carry the evolution, with $\Cov(f_t,h_t)=\frac{\hbar}{2}g_{\rm sol}(X_{f_t},X_{h_t})$. The vacuum is a fixed metric, and evolution moves the Hamiltonian vector fields.

\subsection{Metric-preserving and squeezing parts of the generator}

A fixed reference $g_0$ splits the generator into parts that are skew- and self-adjoint with respect to $g_0$,
\begin{equation}
 F^{\dagger_0}:=g_0^{-1}F^Tg_0=J_0FJ_0,
 \qquad F_{\rm rot}:=\tfrac12(F-F^{\dagger_0}),
 \qquad F_{\rm sq}:=\tfrac12(F+F^{\dagger_0}),
 \label{eq:cartan}
\end{equation}
where the middle equality uses $F^T\Om+\Om F=0$. Then $[F_{\rm rot},J_0]=0$ and $F_{\rm sq}J_0+J_0F_{\rm sq}=0$. The first part generates the stabilizer $U(N)$ of the reference; the second has eigenvalues in pairs{\footnote{If $u$ is an eigenvector of $F_{\rm sq}$ with eigenvalue $+\sigma$, then the $g_{0}$-orthogonal vector $J_{0}u$ is an eigenvector with eigenvalue $-\sigma $.}} $\pm\sigma_j$ and moves the reference vacuum. This is the metric form of the split of a quadratic Hamiltonian into number-preserving and pair-creation terms. At an instant when $g=g_0$, Eq.~\eqref{eq:evolution} gives $\dot g=-2g_0F_{\rm sq}$.

For one canonical pair the split is explicit. In coordinates $(Q,P)$ adapted to the reference, so that $g_0=\hbar(\dd Q^2+\dd P^2)$, write
\begin{equation}
 F_{\rm rot}=\Om_{\rm r}\begin{pmatrix}0&1\\-1&0\end{pmatrix},\qquad F_{\rm sq}=\begin{pmatrix}s_1&s_2\\s_2&-s_1\end{pmatrix},\qquad \sigma=\sqrt{s_1^2+s_2^2}.
 \label{eq:onepairsplit}
\end{equation}
The annihilation variable $a=(Q+iP)/\sqrt2$ of the reference then obeys
\begin{equation}
 \dot a=-i\Om_{\rm r}\,a+\zeta\,\bar a,\qquad \zeta=s_1+is_2,\quad|\zeta|=\sigma .
 \label{eq:complexgen}
\end{equation}
The rotation part multiplies $a$ by a phase and the squeezing part mixes $a$ with $\bar a$. If $a(t)=\alpha(t)a(t_0)+\beta(t)\bar a(t_0)$, the Bogoliubov coefficients satisfy $i\dot\beta=\Om_{\rm r}\beta+i\zeta\bar\alpha$. This is the evolution equation of the squeezing formalism \cite[Eq.~(3.28)]{gv}, $i\dot\beta_k=F_k\beta_k+R_ke^{i\Theta_k}\alpha_k^*$, with $F_k=\Om_{\rm r}$ the rotation rate, $R_k=\sigma$ the positive eigenvalue of $F_{\rm sq}$, and $\Theta_k$ the direction of its principal axis.

\subsection{The rate of squeezing}

For a fixed reference, Eq.~\eqref{eq:evolution} gives the evolution of $L_t=g_0^{-1}g_t$:
\begin{equation}
 \dot L_{t}=-\bigl(F^{\dagger_0}L_{t}+L_{t}F\bigr).
 \label{eq:Ldot}
\end{equation}

\begin{proposition}
\label{prop:rate}
Let $e^{2r(t)}$ be a simple eigenvalue of $L_t$ with eigenvector $u_t$. Then
\begin{equation}
 \frac{\dd r}{\dd t}=-\frac{g_0(u_t,F_{\rm sq}u_t)}{g_0(u_t,u_t)}.
 \label{eq:rate}
\end{equation}
Consequently $|\dot r|\le\sigma_{\max}(t)$, the largest eigenvalue of $F_{\rm sq}(t)$, and $r_{\max}(t)\le r_{\max}(t_0)+\int_{t_0}^t\sigma_{\max}(t')\,\dd t'$.
\end{proposition}

\begin{proof}
Since $L$ is $g_0$-self-adjoint, first-order perturbation theory gives $\dot\lambda=g_0(u,\dot Lu)/g_0(u,u)$ for $\lambda=e^{2r}$. With Eq.~\eqref{eq:Ldot}, $Lu=\lambda u$ and $g_0(u,F^{\dagger_0}v)=g_0(Fu,v)$, this is $\dot\lambda=-2\lambda\,g_0(u,F_{\rm sq}u)/g_0(u,u)$, which is Eq.~\eqref{eq:rate}. The bound follows because the right-hand side is a Rayleigh quotient of $F_{\rm sq}$. The largest eigenvalue of $L$ is the maximum of its Rayleigh quotient, hence Lipschitz in $t$, and the same estimate holds for it almost everywhere.
\end{proof}

The metric-preserving part $F_{\rm rot}$ drops out entirely: it rotates the principal directions of the state but cannot change its squeezing. The bound is saturated only when the principal axis of $L$ is aligned with that of $F_{\rm sq}$; Section~\ref{sec:twocharts} gives an example of each case. For the constant-frequency oscillator $F_{\rm sq}=0$ in reference-adapted coordinates, and an initially squeezed vacuum rotates at constant $r$: ordinary oscillator evolution generates exactly the rotation of Section~\ref{sec:rotation}.

\subsection{Reference choices}

For a time-dependent reference one compares $L_t=g_{\rm ref}(t)^{-1}g_{\rm state}(t)$. Transforming both metrics by the same canonical map preserves its spectrum; transporting the reference with the state keeps $L_t=\Id$; selecting an instantaneous or adiabatic reference is a separate prescription and generally gives a different comparison. Particle production and quadrature squeezing are meaningful only once the reference has been specified; they cannot be read off the appearance of a time-dependent matrix.

\section{Cosmology: a fixed vacuum and moving observables}
\label{sec:cosmology}

\subsection{The mode and its covariance}
\label{sec:modecov}

Consider a massless minimally coupled field in a spatially flat FLRW background, with conformal time $\eta$, on a spatial torus so that the wave vectors are discrete. Expand it in Fourier modes, $\phi(\vec x)\propto\sum_{\bm k}\phi_{\bm k}e^{i\bm k\cdot\vec x}$, with the spatial normalization absorbed into the mode variables. Since the field is real, $\phi_{-\bm k}=\bar\phi_{\bm k}$, so the modes $\bm k$ and $-\bm k$ are not independent. For each such pair we write
\begin{equation}
 \phi_{\bm k}=\frac{\varphi_c+i\varphi_s}{\sqrt2},
 \label{eq:realsectors}
\end{equation}
with $\varphi_c$ and $\varphi_s$ \emph{real}: up to sign and normalization they are the amplitudes of the standing waves $\cos(\bm k\cdot\vec x)$ and $\sin(\bm k\cdot\vec x)$. The action splits into two decoupled real oscillators, each with Lagrangian $\frac12a^2(\varphi'^2-k^2\varphi^2)$, momentum $p_\varphi=a^2\varphi'$, wave number $k=|\bm k|>0$ and equation of motion
\begin{equation}
 \varphi''+2\frac{a'}a\varphi'+k^2\varphi=0 .
\end{equation}
\emph{Throughout Sections~\ref{sec:modecov}--\ref{sec:twocharts}, $\varphi$ denotes one of these real sectors}, $\varphi_c$ or $\varphi_s$, and $p_\varphi$ its momentum. All the phase-space objects below (covariances, canonical charts, references and squeezing parameters) are real and refer to this single canonical pair. Both sectors have the same dynamics and the same mode function. If $\hat a_c$ and $\hat a_s$ are their annihilation operators, the operators of the complex modes are
\begin{equation}
 \hat A_{\bm k}=\frac{\hat a_c+i\hat a_s}{\sqrt2},\qquad \hat A_{-\bm k}=\frac{\hat a_c-i\hat a_s}{\sqrt2},
 \label{eq:complexops}
\end{equation}
so that $\hat\phi_{\bm k}=\sqrt\hbar\,(e_k\hat A_{\bm k}+\bar e_k\hat A^\dagger_{-\bm k})$ with the mode function $e_k$ introduced below. Since this is a fixed linear combination, any vacuum defined sector by sector is also the vacuum of the corresponding complex operators. A state that is squeezed by $r$, with the same orientation, in both real sectors is a two-mode squeezed vacuum with the same parameter $r$ in the pair $(\hat A_{\bm k},\hat A_{-\bm k})$; we return to this at the end of Section~\ref{sec:twocharts}.

In the Heisenberg picture of Section~\ref{sec:dynamics}, the vacuum is a fixed metric on the space of solutions, and the observables $\varphi(\eta)$ and $p_\varphi(\eta)$ have Hamiltonian vector fields that move with $\eta$; their uncertainties are the $g$-norms of these moving vector fields \cite{ack}. Concretely, let $e_k$ be the mode function, with $a^2(e_k\bar e'_k-e'_k\bar e_k)=i$, the field analogue of Eq.~\eqref{eq:wronsk}. Expanding the real sector as $\hat\varphi=\sqrt\hbar(e_k\hat a+\bar e_k\hat a^\dagger)$, a Hermitian operator with $[\hat\varphi,\hat p_\varphi]=i\hbar$, gives, in its vacuum, the real covariance
\begin{equation}
 V_{\varphi\varphi}=\hbar|e_k|^2,\qquad
 V_{pp}=\hbar a^4|e'_k|^2,\qquad
 C_{\varphi p}=\hbar a^2\operatorname{Re}(e_k\bar e'_k),
 \label{eq:modecov}
\end{equation}
the field analogue of Eq.~\eqref{eq:oscfluct}, and the Wronskian gives $\det V_k=\hbar^2/4$. For de Sitter, $a=-1/(H\eta)$, the Bunch--Davies solution $e_k=\frac{H}{\sqrt{2k^3}}(i-k\eta)e^{-ik\eta}$ yields
\begin{equation}
 V_{\varphi\varphi}=\frac{\hbar H^2}{2k^3}(1+k^2\eta^2),\qquad
 V_{pp}=\frac{\hbar k}{2H^2\eta^2},\qquad
 C_{\varphi p}=\frac{\hbar}{2k\eta}.
 \label{eq:ds}
\end{equation}
As $\eta\to0^-$, $V_{\varphi\varphi}$ freezes while $V_{pp}$ and $|C_{\varphi p}|$ grow, and their cancellation in $\det V_k$ keeps the state pure. As in the oscillator puzzle, the covariance is essential: without it the Wronskian identity would look like a failure of minimum uncertainty. The uncertainty ellipse becomes extremely elongated a few $e$-folds after horizon crossing; how elongated depends on the canonical variables used to draw it.

\subsection{Changing the field variable}

For the rescaled field $\chi=a\varphi$ two canonical momenta are natural,
\begin{equation}
 p_\chi=\chi'=\frac{p_\varphi}{a}+a'\varphi,
 \qquad
 \tilde p_\chi=\frac{p_\varphi}{a}=\chi'-\frac{a'}{a}\chi,
 \label{eq:rescaling}
\end{equation}
given by the symplectic maps $T_\eta=\bigl(\begin{smallmatrix}a&0\\a'&a^{-1}\end{smallmatrix}\bigr)$ and $\tilde T_\eta=\diag(a,a^{-1})$. These are the $v$ and $u$ variables{\footnote{Precisely, $(\chi_{c}+i\chi_{s})/\sqrt{2}=v_{\bm{k}}=u_{\bm{k}}$, $(p_{\chi\, c}+ip_{\chi \, s})/\sqrt{2}=p_{v\,\bm{k}}$, and $(\tilde{p}_{\chi\, c}+i\tilde{p}_{\chi \, s})/\sqrt{2}=p_{u\,\bm{k}}$.}}  of Ref.~\cite[Table~1]{gv}. Covariances transform as $V_\chi=T_\eta V_\varphi T_\eta^T$, and because $T_\eta$ depends on time the generator acquires an inhomogeneous term,
\begin{equation}
 F_\chi=T'_\eta T_\eta^{-1}+T_\eta F_\varphi T_\eta^{-1}.
 \label{eq:Fchi}
\end{equation}
Thus, in the chart $(\chi,p_\chi)$ the Hamiltonian is{\footnote{Recall that for quadratic Hamiltonians,  $H=\tfrac{1}{2}x^{T}{\cal{K}}x$, it holds that $F=\Pi {\cal{K}}$. Since ${\cal{K}}_{\varphi}=\diag(k^{2}a^{2},a^{-2})$, we obtain ${\cal{K}}_{\chi}$  from Eq.~\eqref{eq:Fchi}, and thus the Hamiltonian $H_{\chi}$.}} $H_\chi=\frac12[p_\chi^2+(k^2-a''/a)\chi^2]$. In the chart $(\chi,\tilde p_\chi)$ it is
\begin{equation}
 \tilde H_\chi=\tfrac12\bigl(\tilde p_\chi^2+k^2\chi^2\bigr)+\frac{a'}{a}\,\chi\tilde p_\chi ,
 \label{eq:squeezeH}
\end{equation}
with an explicit squeezing term. For de Sitter, $a''/a=2/\eta^2$, and for $|k\eta|\le\sqrt2$ the instantaneous Hamiltonian $H_\chi$ is not positive: it describes a free particle at $|k\eta|=\sqrt2$ and an inverted oscillator below. No late-time reference can be obtained by declaring an instantaneous ground state in these variables; a reference has to be supplied by other physical input. References that minimize the squeezing part of the generator remain available in this regime and are studied in Ref.~\cite{ls}.

\subsection{Two references and their squeezing}
\label{sec:twocharts}

In each chart, consider the vacuum of the fixed-frequency oscillator $\frac12(p^2+k^2\chi^2)$, with $p=p_\chi$ or $p=\tilde p_\chi$:
\begin{equation}
 g_{\rm ref}=k\,\dd\chi^2+k^{-1}\dd p^2,
 \qquad
 V_{\rm ref}=\frac{\hbar}{2}\diag(k^{-1},k).
\end{equation}
At each time the two references differ by a squeeze with $\sinh d_k=aH/(2k)$ \cite[Eq.~(6.9)]{gv}, which in de Sitter equals $1/(2|k\eta|)$. Both references are used in practice.
\begin{itemize}
\item The chart $(\chi,\tilde p_\chi)$, with Hamiltonian~\eqref{eq:squeezeH} and annihilation operators $(k\chi+i\tilde p_\chi)/\sqrt{2\hbar k}$, is the standard squeezing formalism for cosmological perturbations \cite{afjp,mv}, with $\chi$ playing the role of the Mukhanov--Sasaki variable. In the original variables its reference is $g=a^2k\,\dd\varphi^2+(a^2k)^{-1}\dd p_\varphi^2$, the instantaneous ground state of the field Hamiltonian $\frac12(p_\varphi^2/a^2+a^2k^2\varphi^2)$, which is positive at all times \cite{ls,prd}. Those references write it for the complex variable $\chi_{\bm k}$ and its annihilation operators $\hat c_{\pm\bm k}$. By Eq.~\eqref{eq:complexops}, these are fixed combinations of the real-sector operators $(k\chi+i\tilde p_\chi)/\sqrt{2\hbar k}$ of $\chi_c$ and $\chi_s$, so they define the same vacuum, and $\tilde r_k$ below is the same squeezing parameter.
\item The reference of the chart $(\chi,p_\chi)$ is the vacuum of a subsequent radiation era. If inflation ends at $\eta_e$ and is followed by radiation domination with $a$ and $a'$ continuous, then $a''=0$ afterwards, the radiation-era vacuum has $\chi=e^{-ik\eta}/\sqrt{2k}$, and its covariance in $(\chi,\chi')$ is exactly $V_{\rm ref}$. Since $p_\chi=p_\varphi/a+a'\varphi$ is continuous at $\eta_e$, the comparison at $\eta=\eta_e$ is the comparison of the inherited state with the radiation-era vacuum.
\end{itemize}
For a single canonical pair (one real sector) $\cosh2r=\frac12\Tr L$, which with $\det V=\hbar^2/4$ reads
\begin{equation}
 \cosh2r=\frac{1}{\hbar}\left(k\,V_{\chi\chi}+\frac{V_{pp}}{k}\right).
 \label{eq:cosh2r}
\end{equation}
With $x=k\eta<0$, Eq.~\eqref{eq:ds} gives $V_{\chi\chi}=\hbar(1+x^2)/(2kx^2)$ in both charts, and
\begin{equation}
 V_{p_\chi p_\chi}=\frac{\hbar k}{2}\Bigl(1-\frac1{x^2}+\frac1{x^4}\Bigr),
 \qquad
 V_{\tilde p_\chi\tilde p_\chi}=\frac{\hbar k}{2},
\end{equation}
so that
\begin{equation}
 \sinh r_k=\frac{1}{2k^2\eta^2},
 \qquad
 \sinh \tilde r_k=\frac{1}{2|k\eta|},
 \label{eq:twocharts}
\end{equation}
for the radiation-era and the squeezing-formalism references respectively. The second expression is the squeezing parameter usually quoted for de Sitter \cite{afjp,mv}. Both agree with the leading terms of the sub-Hubble expansions of Ref.~\cite[Eqs.~(6.59) and (6.69)]{gv} for a massless field ($\nu=3/2$); here they are exact at all times. The first, evaluated at $\eta_e$, reproduces the sudden-transition occupation number $\sinh^2r_k=1/(4k^4\eta_e^4)=e^{4N_k}/4$ of the radiation era \cite{corichicl}, where $N_k=\ln(1/|k\eta|)$ is the number of $e$-folds since horizon crossing. Figure~\ref{fig:cosmoellipses} draws the Bunch--Davies uncertainty ellipse, at three times, in the coordinates adapted to each reference.

\begin{figure}[t]
\centering
\begin{tikzpicture}[scale=0.62]
\begin{scope}[shift={(0.0,0)}]
\begin{scope}\clip (-4.2,-4.2) rectangle (4.2,4.2);
\draw[gray,dashed] (0,0) circle (1);
\draw[gray!70,very thick] plot[smooth cycle] coordinates {(-0.480,-0.615) (-0.567,-0.544) (-0.648,-0.469) (-0.725,-0.390) (-0.797,-0.309) (-0.862,-0.225) (-0.921,-0.139) (-0.973,-0.052) (-1.017,0.035) (-1.054,0.122) (-1.082,0.208) (-1.103,0.293) (-1.115,0.375) (-1.118,0.454) (-1.113,0.530) (-1.099,0.602) (-1.078,0.669) (-1.048,0.732) (-1.010,0.788) (-0.964,0.839) (-0.911,0.883) (-0.851,0.921) (-0.784,0.951) (-0.712,0.974) (-0.634,0.990) (-0.551,0.999) (-0.464,0.999) (-0.374,0.993) (-0.281,0.978) (-0.185,0.956) (-0.089,0.927) (0.009,0.891) (0.106,0.848) (0.203,0.798) (0.298,0.743) (0.391,0.682) (0.480,0.615) (0.567,0.544) (0.648,0.469) (0.725,0.390) (0.797,0.309) (0.862,0.225) (0.921,0.139) (0.973,0.052) (1.017,-0.035) (1.054,-0.122) (1.082,-0.208) (1.103,-0.293) (1.115,-0.375) (1.118,-0.454) (1.113,-0.530) (1.099,-0.602) (1.078,-0.669) (1.048,-0.732) (1.010,-0.788) (0.964,-0.839) (0.911,-0.883) (0.851,-0.921) (0.784,-0.951) (0.712,-0.974) (0.634,-0.990) (0.551,-0.999) (0.464,-0.999) (0.374,-0.993) (0.281,-0.978) (0.185,-0.956) (0.089,-0.927) (-0.009,-0.891) (-0.106,-0.848) (-0.203,-0.798) (-0.298,-0.743) (-0.391,-0.682) (-0.480,-0.615)};
\draw[teal,very thick] plot[smooth cycle] coordinates {(-0.325,-0.526) (-0.444,-0.450) (-0.559,-0.370) (-0.670,-0.288) (-0.776,-0.203) (-0.876,-0.117) (-0.970,-0.030) (-1.056,0.057) (-1.134,0.144) (-1.203,0.230) (-1.263,0.314) (-1.314,0.395) (-1.354,0.474) (-1.385,0.549) (-1.405,0.620) (-1.414,0.686) (-1.412,0.746) (-1.399,0.802) (-1.376,0.851) (-1.343,0.893) (-1.299,0.929) (-1.245,0.958) (-1.182,0.979) (-1.110,0.993) (-1.030,1.000) (-0.941,0.998) (-0.846,0.990) (-0.743,0.973) (-0.636,0.950) (-0.523,0.919) (-0.407,0.881) (-0.287,0.836) (-0.165,0.785) (-0.042,0.728) (0.081,0.665) (0.204,0.598) (0.325,0.526) (0.444,0.450) (0.559,0.370) (0.670,0.288) (0.776,0.203) (0.876,0.117) (0.970,0.030) (1.056,-0.057) (1.134,-0.144) (1.203,-0.230) (1.263,-0.314) (1.314,-0.395) (1.354,-0.474) (1.385,-0.549) (1.405,-0.620) (1.414,-0.686) (1.412,-0.746) (1.399,-0.802) (1.376,-0.851) (1.343,-0.893) (1.299,-0.929) (1.245,-0.958) (1.182,-0.979) (1.110,-0.993) (1.030,-1.000) (0.941,-0.998) (0.846,-0.990) (0.743,-0.973) (0.636,-0.950) (0.523,-0.919) (0.407,-0.881) (0.287,-0.836) (0.165,-0.785) (0.042,-0.728) (-0.081,-0.665) (-0.204,-0.598) (-0.325,-0.526)};
\draw[rust,very thick] plot[smooth cycle] coordinates {(-0.159,-0.383) (-0.352,-0.301) (-0.543,-0.216) (-0.730,-0.131) (-0.912,-0.044) (-1.086,0.044) (-1.252,0.131) (-1.409,0.216) (-1.555,0.301) (-1.689,0.383) (-1.811,0.462) (-1.918,0.537) (-2.011,0.609) (-2.088,0.676) (-2.150,0.737) (-2.195,0.793) (-2.224,0.843) (-2.236,0.887) (-2.230,0.924) (-2.208,0.954) (-2.169,0.976) (-2.113,0.991) (-2.042,0.999) (-1.954,0.999) (-1.852,0.991) (-1.736,0.976) (-1.607,0.954) (-1.465,0.924) (-1.312,0.887) (-1.149,0.843) (-0.978,0.793) (-0.799,0.737) (-0.614,0.676) (-0.424,0.609) (-0.231,0.537) (-0.036,0.462) (0.159,0.383) (0.352,0.301) (0.543,0.216) (0.730,0.131) (0.912,0.044) (1.086,-0.044) (1.252,-0.131) (1.409,-0.216) (1.555,-0.301) (1.689,-0.383) (1.811,-0.462) (1.918,-0.537) (2.011,-0.609) (2.088,-0.676) (2.150,-0.737) (2.195,-0.793) (2.224,-0.843) (2.236,-0.887) (2.230,-0.924) (2.208,-0.954) (2.169,-0.976) (2.113,-0.991) (2.042,-0.999) (1.954,-0.999) (1.852,-0.991) (1.736,-0.976) (1.607,-0.954) (1.465,-0.924) (1.312,-0.887) (1.149,-0.843) (0.978,-0.793) (0.799,-0.737) (0.614,-0.676) (0.424,-0.609) (0.231,-0.537) (0.036,-0.462) (-0.159,-0.383)};
\end{scope}
\draw[->,gray] (-4.3,0)--(4.4,0) node[right]{$Q$}; \draw[->,gray] (0,-4.3)--(0,4.4) node[above]{$P$};
\node[font=\small] at (0,-5.0) {squeezing-formalism reference};
\end{scope}
\begin{scope}[shift={(9.0,0)}]
\begin{scope}\clip (-4.2,-4.2) rectangle (4.2,4.2);
\draw[gray,dashed] (0,0) circle (1);
\draw[gray!70,very thick] plot[smooth cycle] coordinates {(0.227,-0.853) (0.130,-0.875) (0.033,-0.891) (-0.065,-0.899) (-0.162,-0.901) (-0.257,-0.896) (-0.351,-0.884) (-0.442,-0.866) (-0.530,-0.840) (-0.614,-0.809) (-0.693,-0.771) (-0.767,-0.728) (-0.835,-0.678) (-0.896,-0.624) (-0.951,-0.565) (-0.999,-0.502) (-1.039,-0.434) (-1.071,-0.364) (-1.095,-0.291) (-1.110,-0.215) (-1.118,-0.138) (-1.116,-0.060) (-1.106,0.019) (-1.088,0.097) (-1.061,0.175) (-1.027,0.251) (-0.984,0.326) (-0.934,0.398) (-0.877,0.467) (-0.814,0.532) (-0.744,0.594) (-0.668,0.650) (-0.587,0.702) (-0.502,0.749) (-0.413,0.790) (-0.321,0.825) (-0.227,0.853) (-0.130,0.875) (-0.033,0.891) (0.065,0.899) (0.162,0.901) (0.257,0.896) (0.351,0.884) (0.442,0.866) (0.530,0.840) (0.614,0.809) (0.693,0.771) (0.767,0.728) (0.835,0.678) (0.896,0.624) (0.951,0.565) (0.999,0.502) (1.039,0.434) (1.071,0.364) (1.095,0.291) (1.110,0.215) (1.118,0.138) (1.116,0.060) (1.106,-0.019) (1.088,-0.097) (1.061,-0.175) (1.027,-0.251) (0.984,-0.326) (0.934,-0.398) (0.877,-0.467) (0.814,-0.532) (0.744,-0.594) (0.668,-0.650) (0.587,-0.702) (0.502,-0.749) (0.413,-0.790) (0.321,-0.825) (0.227,-0.853)};
\draw[teal,very thick] plot[smooth cycle] coordinates {(0.325,-0.526) (0.204,-0.598) (0.081,-0.665) (-0.042,-0.728) (-0.165,-0.785) (-0.287,-0.836) (-0.407,-0.881) (-0.523,-0.919) (-0.636,-0.950) (-0.743,-0.973) (-0.846,-0.990) (-0.941,-0.998) (-1.030,-1.000) (-1.110,-0.993) (-1.182,-0.979) (-1.245,-0.958) (-1.299,-0.929) (-1.343,-0.893) (-1.376,-0.851) (-1.399,-0.802) (-1.412,-0.746) (-1.414,-0.686) (-1.405,-0.620) (-1.385,-0.549) (-1.354,-0.474) (-1.314,-0.395) (-1.263,-0.314) (-1.203,-0.230) (-1.134,-0.144) (-1.056,-0.057) (-0.970,0.030) (-0.876,0.117) (-0.776,0.203) (-0.670,0.288) (-0.559,0.370) (-0.444,0.450) (-0.325,0.526) (-0.204,0.598) (-0.081,0.665) (0.042,0.728) (0.165,0.785) (0.287,0.836) (0.407,0.881) (0.523,0.919) (0.636,0.950) (0.743,0.973) (0.846,0.990) (0.941,0.998) (1.030,1.000) (1.110,0.993) (1.182,0.979) (1.245,0.958) (1.299,0.929) (1.343,0.893) (1.376,0.851) (1.399,0.802) (1.412,0.746) (1.414,0.686) (1.405,0.620) (1.385,0.549) (1.354,0.474) (1.314,0.395) (1.263,0.314) (1.203,0.230) (1.134,0.144) (1.056,0.057) (0.970,-0.030) (0.876,-0.117) (0.776,-0.203) (0.670,-0.288) (0.559,-0.370) (0.444,-0.450) (0.325,-0.526)};
\draw[rust,very thick] plot[smooth cycle] coordinates {(-0.201,0.124) (-0.006,0.438) (0.189,0.748) (0.382,1.053) (0.573,1.349) (0.759,1.635) (0.940,1.909) (1.113,2.168) (1.278,2.411) (1.433,2.636) (1.577,2.840) (1.709,3.023) (1.828,3.183) (1.934,3.318) (2.024,3.429) (2.099,3.513) (2.158,3.570) (2.201,3.601) (2.227,3.603) (2.236,3.579) (2.228,3.527) (2.203,3.449) (2.161,3.344) (2.103,3.213) (2.029,3.059) (1.939,2.881) (1.835,2.681) (1.717,2.460) (1.585,2.221) (1.442,1.965) (1.287,1.694) (1.123,1.410) (0.950,1.116) (0.770,0.813) (0.584,0.504) (0.394,0.190) (0.201,-0.124) (0.006,-0.438) (-0.189,-0.748) (-0.382,-1.053) (-0.573,-1.349) (-0.759,-1.635) (-0.940,-1.909) (-1.113,-2.168) (-1.278,-2.411) (-1.433,-2.636) (-1.577,-2.840) (-1.709,-3.023) (-1.828,-3.183) (-1.934,-3.318) (-2.024,-3.429) (-2.099,-3.513) (-2.158,-3.570) (-2.201,-3.601) (-2.227,-3.603) (-2.236,-3.579) (-2.228,-3.527) (-2.203,-3.449) (-2.161,-3.344) (-2.103,-3.213) (-2.029,-3.059) (-1.939,-2.881) (-1.835,-2.681) (-1.717,-2.460) (-1.585,-2.221) (-1.442,-1.965) (-1.287,-1.694) (-1.123,-1.410) (-0.950,-1.116) (-0.770,-0.813) (-0.584,-0.504) (-0.394,-0.190) (-0.201,0.124)};
\end{scope}
\draw[->,gray] (-4.3,0)--(4.4,0) node[right]{$Q$}; \draw[->,gray] (0,-4.3)--(0,4.4) node[above]{$P$};
\node[font=\small] at (0,-5.0) {radiation-era reference};
\end{scope}
\end{tikzpicture}
\caption{The Bunch--Davies vacuum drawn against each reference, in the reference-adapted coordinates $Q=\sqrt k\,\chi$, $P=p/\sqrt k$ (contours $x^TV^{-1}x=2$, $\hbar=1$). Gray, teal and red: $|k\eta|=2$, $1$ and $1/2$. The dashed circle is the reference vacuum. Left: $\tilde r_k=0.25$, $0.48$, $0.88$. Right: $r_k=0.12$, $0.48$, $1.44$. The state is the same in both panels; only the reference differs. The two references happen to give equal squeezing at $|k\eta|=1$.}
\label{fig:cosmoellipses}
\end{figure}
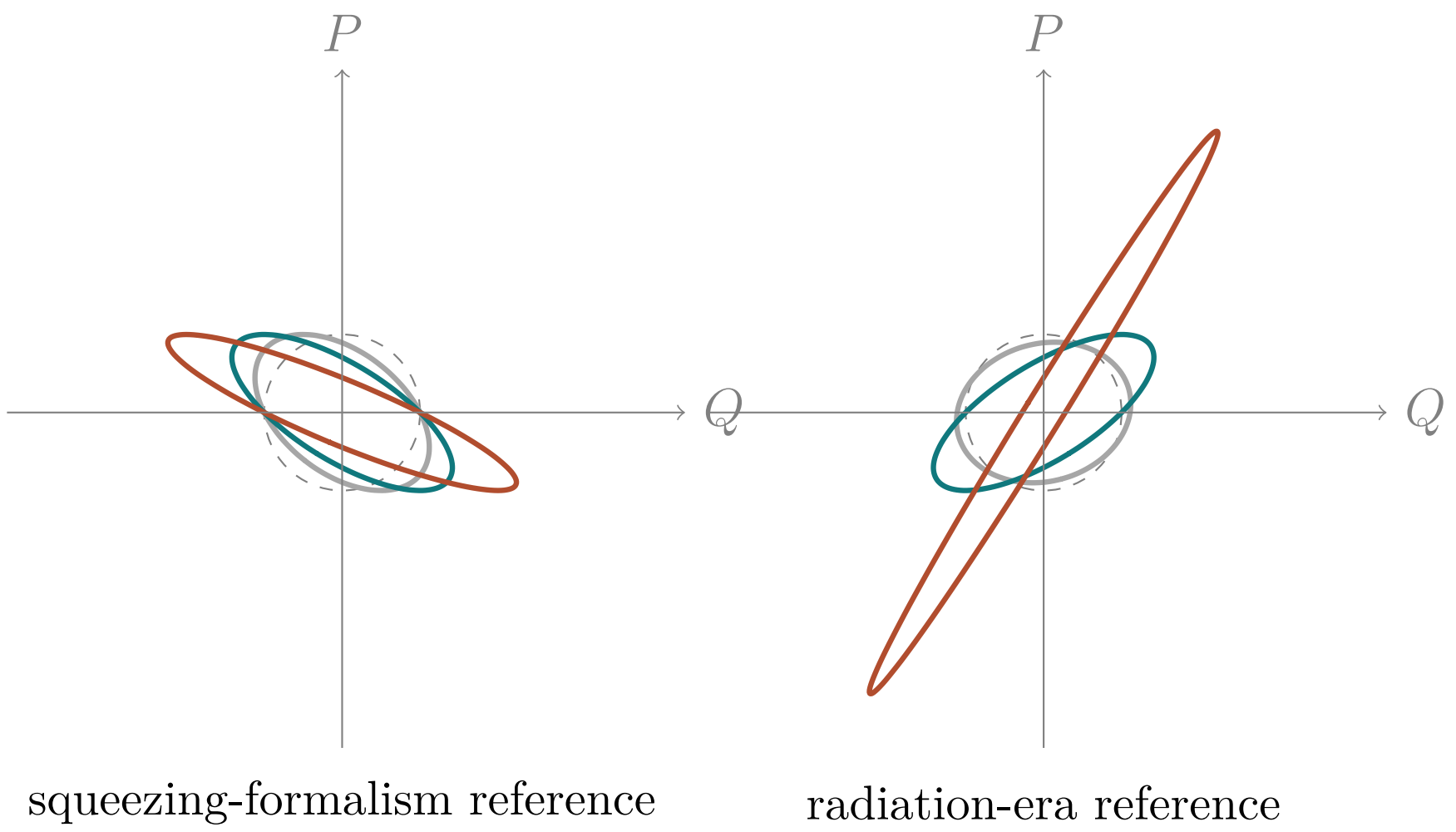

\begin{figure}[t]
\centering
\begin{tikzpicture}
\begin{axis}[
 width=0.62\textwidth, height=6.4cm,
 xmode=log, x dir=reverse,
 xmin=0.01, xmax=10, ymin=0, ymax=10,
 xlabel={$|k\eta|$}, ylabel={relative squeezing},
 legend style={at={(1.03,1)},anchor=north west,font=\small,draw=none,fill=none},legend cell align=left,
 tick label style={font=\small}, label style={font=\small},
 samples=160, domain=-2:1]
\addplot[deepblue,very thick] ({10^x},{ln(1/(2*(10^x)^2)+sqrt(1/(4*(10^x)^4)+1))});
\addlegendentry{$r_k$, radiation-era reference}
\addplot[rust,very thick] ({10^x},{ln(1/(2*10^x)+sqrt(1/(4*(10^x)^2)+1))});
\addlegendentry{$\tilde r_k$, squeezing formalism}
\addplot[black,thin,dashed,domain=-2:0] ({10^x},{-2*ln(10^x)});
\addlegendentry{asymptotes $2N_k$, $N_k$}
\addplot[black,thin,dashed,domain=-2:0,forget plot] ({10^x},{-ln(10^x)});
\draw[gray,dotted,thick] (axis cs:1.41421,0) -- (axis cs:1.41421,10);
\node[font=\footnotesize,gray,align=left,anchor=west] at (axis cs:1.3,4.6) {$H_\chi$ not\\positive};
\end{axis}
\end{tikzpicture}
\caption{Relative squeezing of the same Bunch--Davies state against the two references of Section~\ref{sec:twocharts}, Eq.~\eqref{eq:twocharts}. Time runs to the right. Dashed lines are the late-time asymptotes $2N_k$ and $N_k$. To the right of the dotted line $H_\chi$ has no ground state, but both references remain well defined.}
\label{fig:twocharts}
\end{figure}
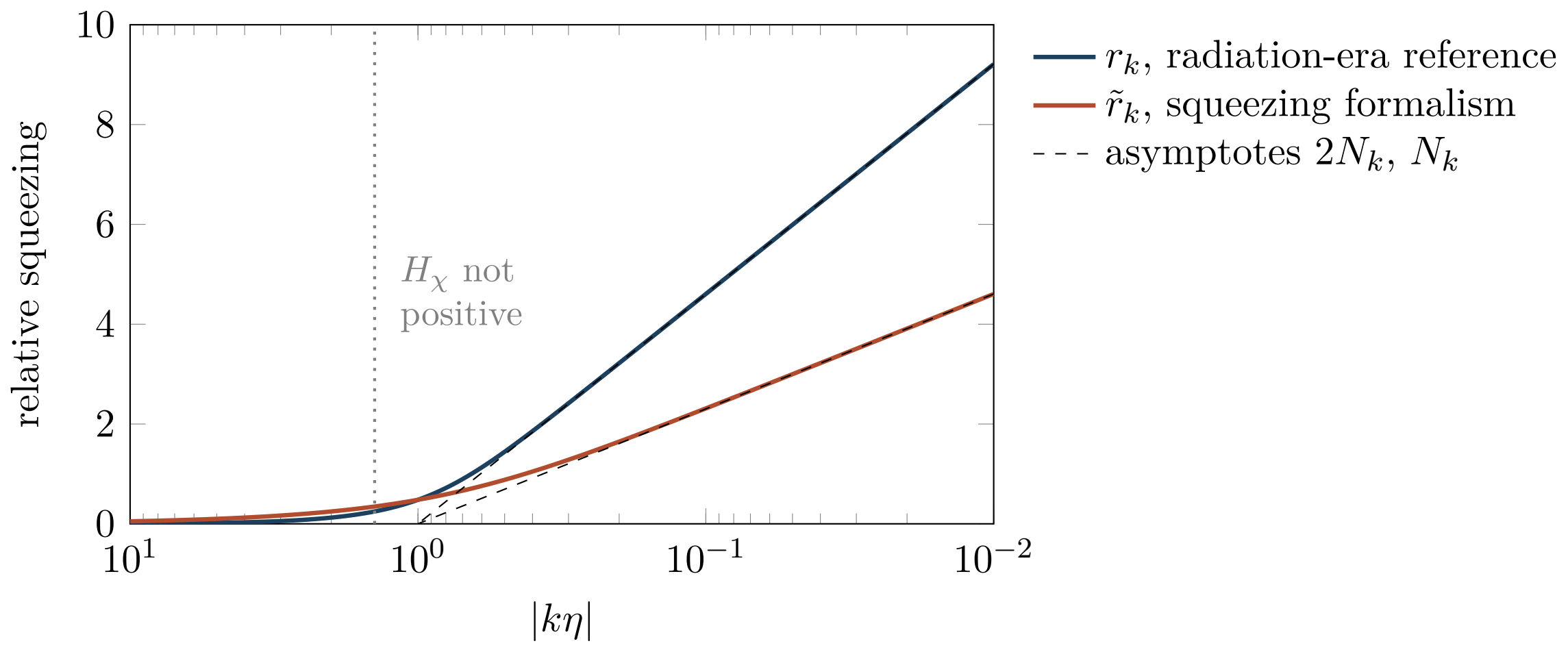

Both parameters vanish as $|k\eta|\to\infty$, where the Bunch--Davies state is the adiabatic vacuum, and after horizon crossing
\begin{equation}
 r_k\simeq2N_k,\qquad \tilde r_k\simeq N_k\qquad(|k\eta|\ll1).
\end{equation}
The statement that the squeezing parameter grows by one unit per $e$-fold therefore refers to the squeezing-formalism reference; relative to the vacuum that a subsequent radiation era selects, the same state squeezes twice as fast (Figure~\ref{fig:twocharts}). Nothing about the state differs between the two columns of Eq.~\eqref{eq:twocharts}; only the reference does. That different canonical variables select different vacua, and hence different predictions, was established in Ref.~\cite{gv}. The relative metric adds an invariant form for the comparison, the late-time rates, and the physical reading of the second reference.

Proposition~\ref{prop:rate} shows where the two rates come from. In reference-adapted coordinates $Q=\sqrt k\,\chi$, $P=p/\sqrt k$, the inhomogeneous term of Eq.~\eqref{eq:Fchi} gives in the chart $(\chi,\tilde p_\chi)$ the generator $F=\bigl(\begin{smallmatrix}a'/a&k\\-k&-a'/a\end{smallmatrix}\bigr)$, whose squeezing part is $F_{\rm sq}=\diag(a'/a,-a'/a)$, with $\sigma=1/|\eta|$. Since $\dd N_k/\dd\eta=1/|\eta|$, the bound reads $\dd\tilde r_k/\dd N_k\le1$, and it is saturated at late times, where the principal axis of $L$ aligns with that of $F_{\rm sq}$. In the chart $(\chi,p_\chi)$ the generator is $F=\bigl(\begin{smallmatrix}0&k\\-\omega^2/k&0\end{smallmatrix}\bigr)$ with $\omega^2=k^2-2/\eta^2$, so $F_{\rm sq}$ is given by $\frac{1}{k\eta^{2}}\bigl(\begin{smallmatrix}0&1 \\ 1&0\end{smallmatrix}\bigr)$ and hence $\sigma=1/(k\eta^2)$. The bound now gives only $\dd r_k/\dd N_k\le1/|k\eta|=e^{N_k}$, far from saturation, because the principal axes of $L$ and $F_{\rm sq}$ are not aligned. The exact rate~\eqref{eq:rate} evaluated on the eigenvector of $L$ gives $\dd r_k/\dd N_k\to2$. The exact rates also follow directly from Eq.~\eqref{eq:twocharts}: both parameters satisfy $\sinh s_{k}=\tfrac{1}{2}e^{\alpha N_{k}}$, with $\alpha=1$ for $\tilde r_k$ and $\alpha=2$ for $r_k$, so that $\dd s_k/\dd N_k=\alpha\tanh s_{k}\leq \alpha$, which tends to $\alpha$ at late times.

All of the above refers to one real sector. The two sectors $\varphi_c$ and $\varphi_s$ of a pair have identical dynamics and identical references, so the Bunch--Davies state is squeezed by the same $r_k$ (or $\tilde r_k$), with the same orientation, in both. By Eq.~\eqref{eq:complexops}, in the complex basis the pair $\pm\bm k$ is then in a two-mode squeezed vacuum with the same parameter. The reduced state of one complex mode is thermal with occupation $\sinh^2r_k$ \cite{corichicl}, and its entanglement entropy, $\simeq2r_k+1-\ln4$, inherits the factor of two between the references. The squeezing parameters are thus the same in the real and the complex description; only the form of the state differs, single-mode squeezing in each standing wave versus two-mode squeezing between $\bm k$ and $-\bm k$. This entanglement is between two global Fourier modes, which local observers cannot access. Between field modes of compact support the Bunch--Davies state behaves differently: compared with the Minkowski vacuum it carries more correlation but less entanglement, and the entanglement decreases as the Hubble rate grows \cite{rab,rabl}.

\section{The space of vacua}
\label{sec:rays}

\subsection{The metric on the Gaussian family}

Compatible metrics label centered pure Gaussian rays, and this family inherits the Fubini--Study metric~\eqref{eq:FS}. For one pair, Eq.~\eqref{eq:wavefunction} provides the expression for the vacua $\psi_{A,B}$ associated with the compatible metrics. Changes in the space of metrics, and consequently in the space of vacua, correspond to variations along $A$ and $B$; the FS metric between the rays $|\psi\rangle$ and $|\psi'\rangle=|\psi + \dd \psi\rangle$ is then of the form $\dd s_{\rm FS}^2=h_{ij}\,\dd y^{i}\dd y^{j}$, with $y^{1}=A$ and $y^{2}=B$, and defines a metric on the space of compatible metrics. To perform the analysis in a concrete manner, it is useful to (re-)parameterize the compatible metrics with a new set $(u,v)$, where $u:=-B^{-1}$ and $v:=A/B$, which decouples the variables in the exponent, yielding the simple form $\psi_{u,v}(q)=(u/{\pi\hbar})^{1/4}\exp[-(u+iv)q^2/{2\hbar}]$ for the vacua. From $\psi_{u,v}$ with $\hbar=1$, Gaussian integration gives
\begin{equation}
 \dd s_{\rm FS}^2=\frac{\dd u^2+\dd v^2}{8u^2}\, .
 \label{eq:FStrace-0}
\end{equation}
On the other hand, compatible metrics [see Eq.~\eqref{eq:oscmetric}] take the form $K(u,v)= u^{-1}\bigl(\begin{smallmatrix}u^{2}+v^2&v\\v&1\end{smallmatrix}\bigr)$ in the $(u,v)$ parametrization. A straightforward calculation shows that the trace of $(K^{-1}\dd K)^{2}$ equals $2(\dd u^{2}+\dd v^{2})/u^{2}$, so we obtain
\begin{equation}
 \dd s_{\rm FS}^2=\frac{\dd u^2+\dd v^2}{8u^2}=\frac1{16}\Tr\bigl[(g^{-1}\dd g)^2\bigr],
 \label{eq:FStrace}
\end{equation}
where $g=\hbar K(u,v)$. The trace form holds for $N$ pairs and is the Gaussian case of the quantum Fisher metric \cite{monras,safranek}. In squeezing coordinates $K=R(\theta)\diag(e^{-2r},e^{2r})R(\theta)^T$,
\begin{equation}
 \dd s_{\rm FS}^2=\frac12\bigl[\dd r^2+\sinh^2(2r)\,\dd\theta^2\bigr]
 =\frac{|\dd z|^2}{2(1-|z|^2)^2},\qquad z=e^{2i\theta}\tanh r,
 \label{eq:FSsqueeze}
\end{equation}
a hyperbolic disk of curvature $-8$ (Figure~\ref{fig:disk}). At $r=0$ the angle is redundant, because rotating an isotropic vacuum does not change its ray. Symplectic maps act on the disk by isometries. For evolution starting{\footnote{In that case, as we have seen before, $\dot{g}=-2g_{0}F_{\rm sq}$.}} at $g=g_0$,
\begin{equation}
 \left(\frac{\dd s_{\rm FS}}{\dd t}\right)^2=\frac14\Tr(F_{\rm sq}^2),
 \label{eq:speed}
\end{equation}
which vanishes for the metric-preserving part of the generator and equals the projective quantum speed $(\Delta H)^2/\hbar^2$ of Eq.~\eqref{eq:AAspeed} in the initial vacuum.

\begin{proposition}
\label{prop:orthogonal}
On the family of Gaussian states $|\gamma;g\rangle$, labelled by a displacement $\gamma$ and a compatible metric $g$, displacements are Fubini--Study orthogonal to deformations of $g$, and
\begin{equation}
 \dd s_{\rm FS}^2=\frac{1}{2\hbar}\,g(\dd\gamma,\dd\gamma)+\frac1{16}\Tr\bigl[(g^{-1}\dd g)^2\bigr].
 \label{eq:FSfull}
\end{equation}
\end{proposition}

\begin{proof}
The first term is Proposition~\ref{prop:coherent}. A variation of $g$ at fixed $\gamma$ changes the state by a centered quadratic operator, whose covariance with any centered linear operator vanishes because odd Gaussian moments vanish; hence the cross terms are zero, and the second term is Eq.~\eqref{eq:FStrace}.
\end{proof}

Equation~\eqref{eq:FSfull} completes the parallel of Section~\ref{sec:geometry}. Along displacements, the quantum metric is the phase-space metric $g$; along changes of vacuum, it is the metric of the space of vacua, which is not a phase-space norm. The squeezing parameters measure distances in the second factor.

\subsection{Overlaps and distances}
\label{sec:distances}

The squared overlap of two centered vacua in physical Darboux coordinates is
\begin{equation}
 |\langle0;g_1|0;g_2\rangle|^2=\det\!\left(\frac{g_1+g_2}{2}\right)^{-1/2},
 \label{eq:overlap}
\end{equation}
which follows from $|\langle\psi_1|\psi_2\rangle|^2=(2\pi\hbar)^N\int\rho_1\rho_2\,\dd^{2N}x$ and the Wigner densities~\eqref{eq:wigner}, and (by applying Theorem \ref{thm:classification}) reduces to $\prod_i(\cosh r_i)^{-1}$ in normal form. So, the cosine of the geodesic distance between the two vacua in the ambient (projective) space is nothing but the square root of this latter quantity.

The metric~\eqref{eq:FStrace} is a constant multiple of the invariant metric of the Siegel space $\Sp(2N,\R)/U(N)$ \cite{siegel}, a Riemannian symmetric space of non-compact type, in which geodesics are unique and minimizing. The curve $s\mapsto g(s)=g_0L^s$ consists of compatible metrics and is the geodesic from $g(0)=g_0$ to $g(1)=g$. A direct calculation shows that $g(s)^{-1}\dd g(s)=(\log L)\dd s$; hence, substituting into Eq.~\eqref{eq:FStrace} and integrating, one finds that the intrinsic FS length of the geodesic connecting $g_0$ and $g$ is $\frac14[\Tr(\log L)^2]^{1/2}=(\frac12\sum_ir_i^2)^{1/2}$. This proves item~4 of Corollary~\ref{cor:invariants}. The ambient Fubini--Study distance is shorter in general, because the shortest path between two Gaussian rays leaves the Gaussian family: along a radial single-mode path the intrinsic length is $r/\sqrt2$, while the ambient distance from the reference is $\arccos[(\cosh r)^{-1/2}]$.

\section{Quantum fields: implementability as finite distance}
\label{sec:fields}

For a free field, a compatible complex structure on the space of solutions selects a pure quasifree state \cite{am}. A deformed vacuum is a normalized vector in the reference Fock space only if the corresponding Bogoliubov transformation is unitarily implementable. The relative metric $L$ expresses this condition directly and connects it with the geometry of Section~\ref{sec:rays}.

\begin{theorem}
\label{thm:shale}
Let $J_0$ and $J$ be compatible complex structures on the real symplectic space of solutions, and assume that $L=g_0^{-1}g$ is bounded with bounded inverse, so that $g$ and $g_0$ define equivalent norms. The following are equivalent:
\begin{enumerate}
\item[(a)] the quasifree state of $J$ is a vector state in the Fock representation of $J_0$, equivalently the Bogoliubov transformation relating them is unitarily implementable;
\item[(b)] $J-J_0$ is Hilbert--Schmidt;
\item[(c)] $L-\Id$ is Hilbert--Schmidt.
\end{enumerate}
If, moreover, $L$ has pure point spectrum $\{e^{\pm2r_i}\}$, they are also equivalent to each of
\begin{enumerate}
\item[(d)] finite reference occupation, $\langle\hat N_0\rangle_J=\sum_i\sinh^2r_i=\frac18\Tr(L+L^{-1}-2)<\infty$;
\item[(e)] nonvanishing overlap, $\prod_i(\cosh r_i)^{-1}>0$, the limit of the finite-dimensional overlaps;
\item[(f)] finite intrinsic distance $\bigl(\frac12\sum_ir_i^2\bigr)^{1/2}$ from $g_0$ to $g$ in the Gaussian family.
\end{enumerate}
\end{theorem}

\begin{proof}
(a)$\Leftrightarrow$(b) is Shale's theorem \cite{shale}; see Ref.~\cite{dg} for the formulation in terms of complex structures. (b)$\Leftrightarrow$(c): $L-\Id=-J_0J+J_0^2=-J_0(J-J_0)$, and $J_0$ is bounded with bounded inverse. With pure point spectrum, the eigenvectors $u$ and $J_0u$ of $L$ give the normal form~\eqref{eq:normalform} mode by mode, so (c) reads $\sum_i[(e^{2r_i}-1)^2+(e^{-2r_i}-1)^2]<\infty$, (d) reads $\sum_i\sinh^2r_i<\infty$, (e) reads $\sum_i\ln\cosh r_i<\infty$, and (f) reads $\sum_ir_i^2<\infty$. Since the $r_i$ are bounded, each summand is bounded above and below by constant multiples of $r_i^2$, so all four conditions are equivalent to $\sum_ir_i^2<\infty$.
\end{proof}

A vacuum is thus unitarily implementable relative to a reference exactly when it lies at finite distance from it in the space of Gaussian vacua. Unitarily inequivalent vacua are infinitely far apart, and the distance, the overlap and the particle number are three measures of the same quantity.

\paragraph{Application to the cosmological example.}
Place the field on a spatial torus of side $\ell$, so that $\bm k\in(2\pi/\ell)\mathbb Z^3\setminus\{0\}$ and the infrared end of mode sums is controlled; the zero mode requires separate treatment. Two different questions must be distinguished.

\emph{Comparison at fixed time.} At any fixed $\eta$, compare the Bunch--Davies state with each reference of Section~\ref{sec:twocharts}. By Eq.~\eqref{eq:twocharts},
\begin{equation}
 \sum_{\bm k}\sinh^2r_k=\sum_{\bm k}\frac{1}{4k^4\eta^4}<\infty,
 \qquad
 \sum_{\bm k}\sinh^2\tilde r_k=\sum_{\bm k}\frac{1}{4k^2\eta^2}\, \longrightarrow\, \infty,
\end{equation}
since the number of modes with $|\bm k|<\Lambda$ grows as $\Lambda^3$. Here each lattice point counts one real sector: the pair $\{\bm k,-\bm k\}$ contributes the two standing-wave sectors of Section~\ref{sec:modecov}, so every real oscillator is counted once. The state is a normalizable vector in the Fock space of the radiation-era reference, but at no time in that of the squeezing-formalism reference: the total occupation number of the standard squeezing formalism is ultraviolet divergent. For the second reference this is the zero-overlap result of Ref.~\cite[Eq.~(6.60)]{gv}, obtained there for a vacuum set at a finite initial time; Theorem~\ref{thm:shale} identifies it with infinite particle number and infinite distance.

\emph{Unitarity of the dynamics.} Fix one reference and ask whether evolution between two times $\eta_0<\eta_f$ is implementable on its Fock space. Let $S_k=S_k(\eta_f,\eta_0)$ be the evolution of one real sector in a given chart, and let $r_k^S$ be the squeezing of the evolved reference $S_k\cdot g_{\rm ref}=S_k^{-T}g_{\rm ref}S_k^{-1}$ relative to $g_{\rm ref}$. The Bogoliubov coefficients of the evolution, defined as below Eq.~\eqref{eq:complexgen}, satisfy $|\alpha_k|^2+|\beta_k|^2=\frac12\Tr(S_kS_k^T)=\cosh2r_k^S$ in reference-adapted coordinates, so that $|\beta_k|=\sinh r_k^S$. By Theorem~\ref{thm:shale}, applied to $J=SJ_0S^{-1}$, the evolution is unitarily implementable if and only if $\sum_{\bm k}(r_k^S)^2<\infty$, equivalently $\sum_{\bm k}|\beta_k|^2<\infty$. The following observation reduces this question to the comparison at fixed time.

\begin{proposition}
\label{prop:dynamics}
Let $g_*(\eta)$ be the metric, in the chart variables at time $\eta$, of a state carried by the dynamics, $S_k\cdot g_*(\eta_0)=g_*(\eta_f)$, and let $r_k^*(\eta)$ be its squeezing relative to $g_{\rm ref}$. Then
\begin{equation}
 \bigl|r_k^*(\eta_f)-r_k^*(\eta_0)\bigr|\le r_k^S\le r_k^*(\eta_f)+r_k^*(\eta_0).
 \label{eq:dynbounds}
\end{equation}
\end{proposition}

\begin{proof}
For one canonical pair the relative squeezing is a distance on the space of vacua, a multiple of the hyperbolic distance of Section~\ref{sec:rays}, and symplectic maps are isometries. The distance from $S_k\cdot g_{\rm ref}$ to $g_*(\eta_f)=S_k\cdot g_*(\eta_0)$ is therefore $r_k^*(\eta_0)$. The triangle with vertices $S_k\cdot g_{\rm ref}$, $g_*(\eta_f)$ and $g_{\rm ref}$ has sides $r_k^*(\eta_0)$, $r_k^*(\eta_f)$ and $r_k^S$, and Eq.~\eqref{eq:dynbounds} is the triangle inequality and its reverse.
\end{proof}

Take $g_*$ to be the Bunch--Davies state, for which Eq.~\eqref{eq:twocharts} gives $r_k^*$ exactly. In the chart $(\chi,p_\chi)$ the upper bound is
\begin{equation}
 r_k^S\le\operatorname{arsinh}\frac{1}{2k^2\eta_0^2}+\operatorname{arsinh}\frac{1}{2k^2\eta_f^2}\le\frac{\eta_0^{-2}+\eta_f^{-2}}{2k^2},
\end{equation}
which is square summable, and the evolution is unitarily implementable. In the chart $(\chi,\tilde p_\chi)$ the lower bound is
\begin{equation}
 r_k^S\ge\operatorname{arsinh}\frac{1}{2k|\eta_f|}-\operatorname{arsinh}\frac{1}{2k|\eta_0|}=\frac{1}{2k}\Bigl(\frac1{|\eta_f|}-\frac1{|\eta_0|}\Bigr)\bigl[1+O(k^{-2})\bigr],
\end{equation}
which is not, and the evolution is not implementable. Both statements hold at every $k$, with no approximation.

The leading behaviour between the two bounds follows from Eq.~\eqref{eq:complexgen}. In the rotating frame $\beta=e^{-i\phi}B$, with $\phi=\int\Om_{\rm r}\,\dd\eta$, the equation for the Bogoliubov coefficient becomes $\dot B=\zeta e^{2i\phi}\bar A$ with $|A|\ge1$. To first order $A=1$, and one integration by parts gives $B\simeq[\zeta e^{2i\phi}/2i\Om_{\rm r}]_{\eta_0}^{\eta_f}$. With $\Om_{\rm r}\simeq k$ and a constant phase of $\zeta$,
\begin{equation}
 |\beta_k|^2=\frac{\sigma_f^2+\sigma_0^2-2\sigma_f\sigma_0\cos(2\Delta\phi)}{4k^2}\bigl[1+O(k^{-1})\bigr],\qquad 2\Delta\phi\simeq2k(\eta_f-\eta_0),
 \label{eq:betaleading}
\end{equation}
where $\sigma_f$ and $\sigma_0$ are the eigenvalues of $F_{\rm sq}$ at the two times. In the chart $(\chi,\tilde p_\chi)$, $\sigma=a'/a$ does not depend on $k$ and $|\beta_k|^2\sim k^{-2}$. In the chart $(\chi,p_\chi)$, $\sigma=(a''/a)/2k=1/(k\eta^2)$ and $|\beta_k|^2\sim k^{-4}$. Since $r_k^*\simeq\sigma/2k$ at each time, Eq.~\eqref{eq:betaleading} is the law of cosines for the triangle of Proposition~\ref{prop:dynamics}, whose angle at $g_*(\eta_f)$ is $2\Delta\phi$: a rotation of the phase plane by $\theta$ turns the disk of Figure~\ref{fig:disk} by $2\theta$. These results are consistent with the analysis of unitary dynamics and uniqueness of Fock quantizations in non-stationary spacetimes \cite{cmov}, in which the choice of canonical pair for the rescaled field is decisive. For general FLRW backgrounds the fixed-time distances of adiabatic states from both references are given in Ref.~\cite{ls}, and Proposition~\ref{prop:dynamics} then applies in the same way.

When implementability fails, the new vacuum can still be a state of the same abstract field algebra, represented by no vector in the reference Fock space. Hadamard short-distance behaviour further restricts physically admissible vacua in curved spacetime \cite{kw,waldbook,wald}, and the finite-dimensional Wigner density~\eqref{eq:wigner} has to be replaced by a Gaussian measure on a distributional completion \cite{ccq,ccq04}.

\section{Discussion}
\label{sec:discussion}

\paragraph{What the phase-space metric contributes.}
We asked what physical information the metric $g$ carries. On the space of quantum states, uncertainty is the norm of the vector field an observable generates; on the quantized phase space, the variance of a linear observable is the $g$-norm of its Hamiltonian vector field; and along coherent states the two metrics coincide (Propositions~\ref{prop:coherent} and~\ref{prop:orthogonal}). The symplectic form determines the commutator and the metric the symmetric part of the two-point function, and compatibility ties them together as appropriate to a pure vacuum. The same metric determines the Schr\"odinger wavefunction, the Weyl characteristic function, and a positive phase-space distribution reproducing Weyl-ordered moments. The inverse metric appears in coordinate covariances because observables are covectors, which prevents the frequent misreading of metric coefficients as coordinate variances.

\paragraph{Squeezed states made simple.}
We asked whether the metric helps to understand squeezed states. A squeezed vacuum is the vacuum of another compatible metric, and squeezing is relational: a single vacuum has no invariant shape, a pair has exactly the invariants $r_i$ (Theorem~\ref{thm:classification}), and every coordinate-independent comparison is a function of them. In this picture the three operations that are often conflated become distinct moves in the space of vacua (Figure~\ref{fig:disk}), and the puzzle of Section~\ref{sec:rotation} disappears: a rotation moves the state around the reference at fixed distance, and the apparent loss of minimum uncertainty is a covariance. Theorem~\ref{thm:shale} extends the picture to fields, where the relative distance can be infinite.

\paragraph{Cosmology.}
We asked whether the metric is useful in cosmology. In the Heisenberg picture the Bunch--Davies vacuum is a fixed metric and the field observables move; their uncertainty ellipse becomes highly elongated after horizon crossing while the state stays pure. How much squeezing one attributes to it depends on the reference. This point was also made in Ref.~\cite{abrm}, where the reference is fixed by the symmetries of de Sitter space; in the present language that reference is the Bunch--Davies state itself, relative to which $r_k=0$. The two references of Ref.~\cite{gv}, both in common use, disagree by a factor of two in the late-time squeezing rate, and only one of them yields a finite total particle number and unitary dynamics. The physically appropriate reference is supplied by the subsequent dynamics, here the radiation era that selects its own vacuum, and not by the inflationary background alone. A positive classical distribution reproducing all Weyl-ordered correlations exists throughout \cite{ack,corichicl}; squeezing makes particular relative uncertainties small, but it does not by itself replace operator products by commuting ones or select an outcome.

\paragraph{Vacua and observers.}
Within the pure Gaussian family, the statement that a squeezed state is the vacuum for another observer is exact as a statement about vacuum prescriptions: the state metric supplies annihilation operators whose vacuum is the state, while relative to $J_0$ the same ray has occupation $\sum_i\sinh^2r_i$. An observer interpretation requires more. If $H_0$ and the observables are held fixed, changing $J$ changes the state and raises the energy to $\hbar\omega(\sinh^2r+\frac12)$. If the state is held fixed and other annihilation operators are used, the particle number changes while expectation values of the original observables do not. Neither operation by itself establishes what a detector on a given trajectory registers; that requires a model of the coupling. For finitely many degrees of freedom all regular irreducible representations are unitarily equivalent \cite{vn}, so both descriptions live in one Hilbert space; for fields they need not.

\paragraph{Purity.}
The identification of a metric with a vacuum uses purity. A reduced mode, such as one member of the pair $(\bm k,-\bm k)$, has a mixed covariance with symplectic eigenvalue $(\hbar/2)\cosh2r_k$, and $g=(\hbar/2)V^{-1}$ then violates the compatibility condition~\eqref{eq:matrixmetric}. Compatibility thus marks exactly where the interpretation as a vacuum geometry stops applying. Mixed Gaussian states are described instead by restricted complex structures, with $J^2\le-\Id$ \cite{hb,amnry}. In that language one can locate, for any mode, the degrees of freedom that carry its correlations and its entanglement, its correlation and entanglement partners \cite{amnry}; for compactly supported modes in de Sitter space this has been carried out in Ref.~\cite{rabl}.

\paragraph{Outlook.}
Natural next steps are to treat entanglement as an invariant of a relative vacuum together with a subsystem decomposition, and to relate it to the reference-free analysis of compactly supported modes \cite{rabl}; to develop the references that minimize the squeezing part of the generator, which remain available where instantaneous ground states do not \cite{ls}, and the robustness of cosmological squeezing against changes of state and reference \cite{prd}; and to describe the space of field vacua as an infinite-dimensional Siegel domain, in which Theorem~\ref{thm:shale} identifies the unitarily implementable vacua with those at finite distance.

\section{Conclusions}
\label{sec:conclusions}

Quantization of a linear system gives its classical phase space a metric, and that metric does on phase space what the quantum metric does on the space of states: fluctuations are metric norms of Hamiltonian vector fields, and along coherent states the two metrics are the same. With the symplectic form fixed, a compatible complex structure and a compatible metric carry the same information, which becomes observable through $\Cov_J(f,h)=(\hbar/2)\,g_J(X_f,X_h)$. The same metric fixes the Schr\"odinger vacuum, the Weyl characteristic function and a positive distribution reproducing Weyl-ordered moments.

Squeezed states then have a simple description. A squeezed vacuum is the vacuum of another metric, and squeezing is a property of a pair of metrics, a relative vacuum. Its invariants are the eigenvalues $e^{\pm2r_i}$ of $L=g_0^{-1}g$, which determine particle content, overlaps, extremal variance ratios and distances. Only the part of the generator that is self-adjoint relative to the reference changes them, at the rate given by Proposition~\ref{prop:rate}. For fields, a vacuum is unitarily implementable relative to a reference exactly when it lies at finite distance from it.

In de Sitter space the same Bunch--Davies state has $\sinh\tilde r_k=1/(2|k\eta|)$ relative to the reference of the standard squeezing formalism and $\sinh r_k=1/(2k^2\eta^2)$ relative to the vacuum that a radiation era beginning at $\eta$ would select. The second squeezes twice as fast and is unitarily implementable, both at fixed time and for the dynamics; the first is not. These two references are those of Grain and Vennin \cite{gv}; the phase-space metric gives their comparison an invariant and exact form. The statement that a squeezed state is another vacuum is exact within the pure Gaussian family. Turning it into a statement about observers, fields or subsystems requires, respectively, a physical prescription for the reference, implementability, and mixed-state data, and the metric viewpoint makes each of these requirements explicit.

\section*{Acknowledgments}

 We thank Ivan Agullo for his valuable comments.

\end{document}